\documentclass[aps,pre,twocolumn,superscriptaddress,longbibliography]{revtex4-2}
\usepackage{amsmath}
\usepackage{amsfonts}
\usepackage{amssymb}
\usepackage{lmodern,dsfont}
\usepackage{graphicx}
\usepackage[usenames,dvipsnames]{xcolor}
\usepackage{bm}
\usepackage[english=american]{csquotes}
\usepackage{xr}

\newcommand{\kb}{k_\text{B}}

\newcommand{\av}[1]{\langle #1 \rangle}
\newcommand{\n}{\nonumber}
\newcommand{\nn}{\nonumber \\}

\newcommand{\grad}{\bm{\nabla}}

\renewcommand{\eqref}[1]{Eq.~(\ref{#1})}

\begin{document}

\author{Andreas Dechant}
\affiliation{Department of Physics \#1, Graduate School of Science, Kyoto University, Kyoto 606-8502, Japan}
\author{Shin-ichi Sasa}
\affiliation{Department of Physics \#1, Graduate School of Science, Kyoto University, Kyoto 606-8502, Japan}
\title{Improving thermodynamic bounds using correlations}
\date{\today}

\begin{abstract}
We discuss how to use correlations between different physical observables to improve recently obtained thermodynamics bounds, notably the fluctuation-response inequality (FRI) and the thermodynamic uncertainty relation (TUR).
We show that increasing the number of measured observables will always produce a tighter bound.
This tighter bound becomes particularly useful if one of the observables is a conserved quantity, whose expectation is invariant under a given perturbation of the system.
For the case of the TUR, we show that this applies to any function of the state of the system.
The resulting correlation-TUR takes into account the correlations between a current and a non-current observable, thereby tightening the TUR.
We demonstrate our finding on a model of the $\text{F}_1$-ATPase molecular motor, a Markov jump model consisting of two rings and transport through a two-dimensional channel.
We find that the correlation-TUR is significantly tighter than the TUR and can be close to an equality even far from equilibrium.
\end{abstract}

\maketitle

\section{Introduction}

Entropy production is a fundamental concept of non-equilibrium statistical mechanics.
It relates the asymmetry of microscopic transitions in a system to the measurable loss of energy in the form of heat dissipated into the environment.
For macroscopic systems, measuring the latter thus provides a measure of microscopic time-reversal symmetry breaking.
While the same relation holds for microscopic systems and can be even be formulated on the level of single trajectories \cite{Sek10,Sei12}, measuring the dissipated heat is generally very challenging, as the resulting temperature changes are very small and typically lost among the fluctuations of the noisy environment.
A more practical way to measure the entropy production in microscopic systems is provided by the work of Harada and Sasa \cite{Har05}, who show that the entropy production can be obtained from the violation of the fluctuation-dissipation relation.
We remark that in principle, the entropy production may also be obtained directly from the probabilities of microscopic transitions in the system, however, this requires very good spatial and temporal resolution as well as lots of statistics.

A different way of estimating entropy production has recently been suggested \cite{Li19,Man20,Ots20,Vu20,Hor20} using the thermodynamic uncertainty relation (TUR) \cite{Bar15,Gin16,Dec17,Pie17}.
The TUR establishes a connection between entropy production on the one hand, and measurable currents in the system and their fluctuations on the other hand.
It may be understood as a more precise formulation of the second law, since it not only establishes the positivity of entropy production but also provides a finite lower bound in terms of experimentally accessible quantities.
However, since the TUR is an inequality, there is generally no guarantee that the lower bound is tight, i.~e.~that a useful estimate of entropy production is obtained from a given measurement.
In principle the lower bound can be optimized to produce an accurate estimate of entropy production \cite{Li19,Man20,Ots20,Vu20} and even realize equality \cite{Dec20b}, however, the resulting quantities may not be any easier to measure than the entropy production itself.

From an experimental point of view, it is thus highly desirable to improve the tightness of the bound using available data.
However, the tightness of the bound is also of fundamental interest:
For example, it has been shown \cite{Hwa18} that the TUR is generally not very tight for models of biological molecular motors, with the lower estimate on entropy production being on the order of $10$ to $40 \%$ of the actual value.
This raises the intriguing question of whether evolution is \enquote{bad} at saturating thermodynamic bounds, or whether indeed a tighter bound exists.

So far, applications and extensions of the TUR have mostly focused only on current-like observables (for example the displacement of a particle or the heat exchanged with the environment) \cite{Dec18,Pal20,Liu20}, although it has been found \cite{Koy19,Koy20,Van20} that, in the presence of time-dependent driving, also state-dependent observables (like the instantaneous position or potential energy) may yield information about the entropy production.
While the presence of non-zero average currents clearly distinguishes a non-equilibrium steady state from an equilibrium system; it is thus reasonable that a relation between currents and the entropy production should exist.
By contrast, the average of state-dependent observables is independent of time both in equilibrium and non-equilibrium steady states, intuitively, it seems that such observables can provide no additional information about the steady state entropy production.
As the main result of this article, we show that this intuitive notion is not correct.
We can exploit the correlations between a state-dependent observable $z$ and current $j$ to obtain a tighter version of the TUR.
We formulate the TUR in terms of the transport efficiency $\eta_J$ \cite{Dec17}
\begin{align}
\eta_J = \frac{2 \av{J}^2}{\text{Var}_J \Delta S_\text{irr}} \leq 1 \label{TUR},
\end{align}
where $J$ is the time-integrated current, $\av{J}$ denotes the average and $\text{Var}_J$ the variance, and $\Delta S^\text{irr}$ is the total entropy production.
Our main result is the bound
\begin{align}
\eta_J + {\chi_{J,Z}}^2 \leq 1 ,
\end{align}
where $Z$ is the time-integral of the state-dependent observable $z$ and $\chi_{J,Z} = \text{Cov}_{J,Z}/\sqrt{\text{Var}_J \text{Var}_Z}$, with $\text{Cov}_{J,Z}$ the covariance, is the Pearson correlation coefficient, which satisfies $-1 \leq \chi \leq 1$.
We refer to \eqref{correlation-TUR} as correlation TUR (CTUR).
As a consequence, we obtain a tighter bound on the entropy production
\begin{align}
\frac{\av{J}^2}{\text{Var}_J} \leq \frac{\av{J}^2}{\text{Var}_J \big( 1 - {\chi_{J,Z}}^2 \big)} \leq \frac{1}{2} \Delta S^\text{irr} \label{correlation-TUR},
\end{align}
where the leftmost expression corresponds to the TUR.
Surprisingly, the observable $Z$ can be almost arbitrary, as long as it is the time-integral of a quantity which only depends on the state of the system.
This implies that virtually any additional observable that can be obtained from a measurement may be used to tighten the TUR.
As we show below, a tight bound is generally obtained when $Z$ is chosen as the local average value of $J$.
Importantly, the CTUR can be evaluated using only the experimentally obtained trajectory data and does not require any additional information about the parameters of the model.
Thus suggests that taking into account correlations between observables may indeed be crucial to obtaining accurate estimates of the entropy production in terms of experimentally accessible quantities.

We demonstrate the usefulness of the CTUR by applying it to three distinct examples:
For a model for the $\text{F}_1$-ATPase molecular motor \cite{Zim12,Kaw14}, we find that, while the bound obtained on the entropy production using the TUR for the displacement of the motor is only around $40 \%$ of the actual value, measuring the time-integrated local mean velocity in addition to the displacement and using the CTUR yields an estimate that is about $90 \%$ accurate over a wide range of parameters.
For a Markov jump model, in which two currents are driven through two connected rings, we show that, even though measuring the current in one of the rings can only give an estimate on the contribution to the entropy production stemming from this ring, this estimate can be tightened considerably using the CTUR.
Finally, for transport in a two-dimensional channel, we demonstrate that even simple choices of the state-dependent observable $z$ can yield a significant improvement over the TUR.

\section{Multidimensional FRI and monotonicity of information} \label{sec-monotonic}
The mathematical basis of our results is an extension of the fluctuation-response inequality (FRI) \cite{Dec20} to multiple observables, similar to the multidimensional TUR \cite{Dec18c}.
The FRI gives an upper bound on the ratio $\mathcal{Q}(r)$ between the response of the average of an observable $Y$ to a small perturbation of the system, and its fluctuations in the unperturbed system,
\begin{align}
\frac{\big(\delta \av{Y} \big)^2}{\text{Var}_Y} \leq 2 D_\text{KL}(\tilde{p} \Vert p) .
\end{align}
Here, $\delta \av{Y} = \widetilde{\av{Y}} - \av{Y}$ is the response of the observable $Y$ to the perturbation which changes the probability density describing the system from $p(\omega)$ to $\tilde{p}(\omega)$ and $D_{\text{KL}}(\tilde{p} \Vert p)$ is the Kullback-Leibler divergence between the probability densities.
Here, $\omega$ may be the state of the system, but it may also represent a trajectory of the system during the measurement interval.
When we consider the perturbation to be described by a parameter $\theta$, such that $p(\omega) = p^\theta(\omega)$ and $\tilde{p}(\omega) = p^{\theta + d\theta}(\omega)$, then this is equivalent to the Cram{\'e}r-Rao inequality \cite{Rao45,Cra16}
\begin{align}
\frac{\big(\partial_\theta \av{Y} \big)^2}{\text{Var}_Y} \leq I(\theta) \label{cramer-rao},
\end{align}
where $I(\theta)$ is the Fisher information
\begin{align}
I(\theta) = \int d\omega \ \big(\partial_\theta \ln p^\theta(\omega) \big)^2 p^\theta(\omega) .
\end{align}
With this identification, we can use the Cram{\'e}r-Rao inequality for vector-valued observables, $\bm{Y}^{(K)} = (Y_1,Y_2,\ldots,Y_K)$,
\begin{align}
\mathcal{Q}^{(K)}_Y \equiv \big(\partial_\theta \av{\bm{Y}^{(K)}} \big)^\text{T} \big(\bm{\Xi}_Y^{(K)} \big)^{-1} \big(\partial_\theta \av{\bm{Y}^{(K)}} \big) \leq I(\theta) \label{fri-multi} ,
\end{align}
where the superscript T denotes transposition and $\bm{\Xi}_Y^{(K)}$ is the covariance matrix with entries $(\bm{\Xi}_Y^{(K)})_{i j} = \text{Cov}_{Y_i,Y_j}$.
Note that here we assumed that the observables are not linearly dependent such that the covariance matrix is positive definite.
As noted in Ref.~\cite{Dec18c}, \eqref{fri-multi} is the extension of the FRI to more than one observable.

Next, we want to show that increasing the number of observables results in a tighter bound, i.~e.~that $\mathcal{Q}^{(K)}_Y \leq \mathcal{Q}^{(K+1)}_Y$.
We write the covariance matrix $\bm{\Xi}_Y^{(K+1)}$ of $K+1$ observables as
\begin{align}
&\bm{\Xi}_Y^{(K+1)} = \begin{pmatrix} \bm{A} & \bm{b} \\ \bm{b}^\text{T} & c \end{pmatrix} \quad \text{with} \\
&\bm{A} = \bm{\Xi}_Y^{(K)}, \; b_k = \text{Cov}_{Y_k,Y_{K+1}}, \; c = \text{Var}_{Y_{K+1}} \n .
\end{align} 
We compute its inverse using the block-inversion formula
\begin{align}
&\big(\bm{\Xi}_Y^{(K+1)} \big)^{-1} = \begin{pmatrix} \bm{A}^{-1} & 0 \\ 0 & 0 \end{pmatrix} + \bm{D} \label{cov-inverse}  \\
&\text{with} \quad \bm{D} = \frac{c - \bm{b}^\text{T} \bm{A}^{-1} \bm{b}}{\big(\bm{A}^{-1} \bm{b}\big)^\text{T} \big(\bm{A}^{-1} \bm{b}\big) } \bm{d} \bm{d}^\text{T}, \; \bm{d} = \begin{pmatrix} - (\bm{A}^{-1} \bm{b})^\text{T} \\ 1 \end{pmatrix} \n .
\end{align}
Further, we have the Schur determinant identity
\begin{align}
\det\big(\bm{\Xi}_Y^{(K+1)}\big) = \det\big(\bm{\Xi}_Y^{(K)}\big) \big(c - \bm{b}^\text{T} \bm{A}^{-1} \bm{b} \big)  .
\end{align}
Since $\bm{\Xi}_Y^{(K+1)}$ and $\bm{\Xi}_Y^{(K)}$ are positive definite, the second factor on the right-hand side is also positive.
As a consequence, the matrix $\bm{D}$ in \eqref{cov-inverse} is positive semi-definite and we have for any $(K+1)$-vector $\bm{v}^{(K+1)}$,
\begin{align}
\bm{v}^{(K+1),\text{T}} &\big(\bm{\Xi}_Y^{(K+1)}\big)^{-1} \bm{v}^{(K+1)} \\
 &= \bm{v}^{(K),\text{T}} \big(\bm{\Xi}_Y^{(K)} \big)^{-1} \bm{v}^{(K)} + \bm{v}^{(K+1),\text{T}} \bm{D} \bm{v}^{(K+1)} \nn
&\geq \bm{v}^{(K),\text{T}} \big(\bm{\Xi}_Y^{(K)}\big)^{-1} \bm{v}^{(K)} \n,
\end{align}
where $\bm{v}^{(K)}$ is the vector $\bm{v}^{(K+1)}$ with the $(K+1)$-th component removed.
For $\bm{v}^{(K+1)} = \partial_\theta \av{\bm{Y}^{(K+1)}}$ this yields the desired inequality 
\begin{align}
\mathcal{Q}^{(K+1)}_Y \geq \mathcal{Q}^{(K)}_Y. \label{information-monotonic}
\end{align}
In light of the Cram{\'e}r-Rao inequality \eqref{fri-multi}, this means that considering more observables yields more information about the parameter $\theta$ (i.~e.~the perturbation) and thus a tighter lower bound on the Fisher information.
In that sense, the information obtained from a measurement increases monotonically with increasing the number of measured observables.
This holds true only as long as the additional observables are not linearly dependent on the existing ones; if this is not the case, then the covariance matrix becomes singular and the bound saturates, as the additional observables do not contain any new information.

In the case of two observables $Y_1$ and $Y_2$, the inverse of the covariance matrix can be computed explicitly and we obtain the bound
\begin{widetext}
\begin{align}
\frac{\big(\partial_\theta \av{Y_1} \big)^2 \text{Var}_{Y_2} - 2 \big(\partial_\theta \av{Y_1} \big) \big(\partial_\theta \av{Y_2} \big) \text{Cov}_{Y_1,Y_2} + \big(\partial_\theta \av{Y_2} \big)^2 \text{Var}_{Y_1}}{\text{Var}_{Y_1} \text{Var}_{Y_2} - {\text{Cov}_{Y_1,Y_2}}^2} \leq I(\theta).
\end{align}
\end{widetext}
This expression simplifies further if $Y_2$ is a conserved quantity with respect to the perturbation, $\partial_\theta \av{Y_2} = 0$.
Then, we find
\begin{align}
\frac{\big(\partial_\theta \av{Y_1} \big)^2}{\text{Var}_{Y_1} \big( 1 - {\chi_{Y_1,Y_2}}^2 \big)} \leq I(\theta) \label{fri-corr}.
\end{align}
In this case it is obvious that the bound is tighter than \eqref{cramer-rao}.
This shows that, even if the average of $Y_2$ contains no information about the parameter $\theta$ and the perturbation, we may still use its correlations with $Y_1$ to obtain a tighter version of the Cram{\'e}r-Rao inequality and thus the FRI.

\section{Continuous time-reversal and TUR} \label{sec-CTUR}
In view of later applications, we slightly generalize the discussion to a Langevin dynamics in $\mathbb{R}^N$ with an internal degree of freedom
\begin{align}
\dot{\bm{x}}(t) = \bm{a}_{i(t)}(\bm{x}(t)) + \bm{G}_{i(t)} \bm{\xi}(t) , \label{langevin}
\end{align}
where the drift vector $\bm{a}_i(\bm{x})$ and diffusion matrix $\bm{G}_i$ depend on the discrete state $i = 1, \ldots, M$.
The dynamics of the discrete state are governed by a Markov jump process with transition rates $W_{ij}(\bm{x})$ from state $j$ to state $i$.
We take the diffusion matrix to be independent of the position in order to simplify some of the following notation, however, the extension to a position-dependent diffusion matrix can be readily obtained.
The evolution of the probability density $p_i(\bm{x},t)$ for being at position $\bm{x}$ and in state $i$ at time $t$ is governed by the Fokker-Planck master equation
\begin{align}
\partial_t p_i(\bm{x},t) = -\grad \big( &\bm{\nu}_i(\bm{x},t) p_i(\bm{x},t) \big) \label{fpe}  \\
& + 2 \sum_j V_{i j}(\bm{x},t) p_j(\bm{x},t) \big),  \nn
\text{with} \quad \bm{\nu}_i(\bm{x},t) &= \bm{a}_i(\bm{x}) - \grad^\text{T} \bm{B}_i \ln p_i(\bm{x},t), \nn
\text{and}  \quad V_{i j}(\bm{x},t) &= \frac{1}{2} \bigg(W_{i j}(\bm{x}) - W_{ji}(\bm{x}) \frac{p_i(\bm{x},t)}{p_j(\bm{x},t)} \bigg) \n .
\end{align}
Here, $\bm{B}_i = 2 \bm{G}_i \bm{G}_i^\text{T}$ is assumed to be positive definite (i.~e.~$\bm{G}_i$ should have full rank).
This dynamics reduces to a pure Langevin dynamics in absence of the discrete degree of freedom and to a pure Markov jump dynamics if there is no dependence on $\bm{x}$.
The quantity $\bm{\nu}_i(\bm{x},t)$ is called the local mean velocity and characterizes the local flows in the system. 
We have also introduced its analog $V_{ij}(\bm{x},t)$ for the jump part.
For the type of dynamics \eqref{langevin}, we may consider two flavors of currents \cite{Che15}
\begin{subequations}
\begin{align}
J_\text{d} &= \int_0^\tau dt \ \bm{w}^\text{T}_{i(t)}(\bm{x}(t)) \circ \dot{\bm{x}}(t)  ,\\
J_\text{j} &= \int_0^\tau \ \omega_{j(t+dt), j(t)}(\bm{x}(t)) .
\end{align} \label{current}%
\end{subequations}
Here $\bm{w}_i(\bm{x})$ is a differentiable vector field, $\omega_{i j}(\bm{x}) = - \omega_{j i}(\bm{x})$ are the entries of an antisymmetric matrix and $\circ$ denotes the Stratonovich product.
Intuitively, the diffusive current $J_\text{d}$ may be interpreted as a generalized displacement, in which the velocity is weighted by the position- and state-dependent function $\bm{w}_i(\bm{x})$.
The jump current $J_\text{j}$, on the other hand, counts transitions between different states, which are weighted by the function $\omega_{i j}(\bm{x})$.
The averages of these quantities in the steady state are proportional to time and given by
\begin{subequations}
\begin{align}
\av{J_\text{d}} &= \tau \sum_{i} \int d\bm{x} \ \bm{w}^\text{T}_{i(t)}(\bm{x}(t)) \bm{\nu}^\text{st}_i(\bm{x}) p_i^\text{st}(\bm{x}) , \\
\av{J_\text{j}} &= \tau \sum_{i j} \int d\bm{x} \ \omega_{i j}(\bm{x})  V_{i j}^\text{st}(\bm{x}) p_j^\text{st}(\bm{x}) .
\end{align} \label{current-average}%
\end{subequations}
Here the superscript st denotes the steady state value of the respective quantity.
Note that both types of current are proportional to the respective local mean velocity.
Next, we briefly summarize the continuous time-reversal introduced in Ref.~\cite{Dec20b}.
This transformation is defined by a family of dynamics of the type \eqref{langevin}, with a parameter $\theta \in [-1,1]$; $\theta = 1$ corresponds to the time-forward dynamics, while $\theta = -1$ represents the time-reversed dynamics \cite{Sas14}.
So, we can connect the time-forward and time-reversed dynamics by a continuous family of dynamics.
In the present context, the most important property of this transformation is that it leads to a rescaling of the local mean velocities $\bm{\nu}_i^{\text{st},\theta}(\bm{x}) = \theta \bm{\nu}_i^\text{st}(\bm{x})$ and $V_{ij}^{\text{st},\theta}(\bm{x}) = \theta V_{ij}^{\text{st}}(\bm{x})$, while the steady state probability density $p_i^\text{st}(\bm{x})$ is independent of $\theta$.
This generalizes the intuitive notion that time-reversal should lead to a reversal of all flows in the system to a continuous transformation.
From \eqref{current-average}, we then see that the averages of currents are also rescaled by the continuous time-reversal operation:
\begin{align}
\av{J}^\theta = \theta \av{J} .
\end{align}
This implies that $\partial_\theta \av{J}^\theta = \av{J}$.
Further, the Fisher information corresponding to the path probabilities of the dynamics parameterized by $\theta$ is related to the entropy production \cite{Has19b,Dec20,Dec20b},
\begin{align}
I(\theta) \leq \frac{1}{2} \Delta S^\text{irr},
\end{align}
where equality holds for a pure Langevin dynamics.
With this, \eqref{fri-multi} implies the multidimensional TUR \cite{Dec18c},
\begin{align}
\av{\bm{Y}^{(K)}}^\text{T} \big(\bm{\Xi}_Y^{(K)} \big)^{-1} \av{\bm{Y}^{(K)}} \leq \frac{1}{2} \Delta S^\text{irr} \label{TUR-multi} ,
\end{align}
where the components of $\bm{Y}^{(K)}$ are currents of either type in \eqref{current}.
The new insight from the preceding discussion is that the left-hand side increases monotonically when increasing the number of measured currents.
This fact is very useful when we want to use the left-hand side to estimate the entropy production:
Any additional information that can be obtained from a measurement can be used to improve the estimate.
We note that, in the steady state of \eqref{langevin}, the entropy production is explicitly given by
\begin{align}
\Delta S^\text{irr} &= \tau \bigg( \sum_i \int d\bm{x} \ \big\Vert \big(\bm{a}_i(\bm{x}) - \grad^\text{T} \bm{B}_i \ln p^\text{st}_i(\bm{x}) \big\Vert^2 p_i^\text{st}(\bm{x})  \nn
&+ \sum_{i,j} \int d\bm{x} \ \ln \bigg(\frac{W_{ij}(\bm{x}) p_j^\text{st}(\bm{x})}{W_{ji}(\bm{x}) p_i^\text{st}(\bm{x})} \bigg) W_{ij}(\bm{x}) p_j^\text{st}(\bm{x}). \label{entropy}
\end{align}

Crucially, \eqref{TUR-multi} is not restricted to current observables. 
To see this, we recall that the steady state probability density $p^\text{st}_{i}(\bm{x})$ is invariant under changing the parameter $\theta$ \cite{Dec20b}.
As a consequence, for a state-dependent (or non-current) observable $Z$
\begin{align}
Z = \int_0^\tau dt \ z_{i(t)}(\bm{x}(t),t) \label{nc-observable},
\end{align}
where the function $z_{i}(\bm{x},t)$ may depend on the position, the internal state and time, its average does not depend on $\theta$
\begin{align}
\av{Z}^\theta = \int_0^\tau dt \sum_i \int d\bm{x} \ z_i(\bm{x},t) p_i^\text{st}(\bm{x}) = \av{Z} .
\end{align}
Thus, $\partial_\theta \av{Z}^\theta = 0$, and we may include such observables in \eqref{TUR-multi} by setting the corresponding entries in the vector $\av{\bm{Y}^{(K)}}$ to zero.
However, such observables do contribute to the covariance matrix $\bm{\Xi}_Y^{(K)}$, and \eqref{information-monotonic} guarantees that the resulting bound will be tighter than the one without these observables.
For the case of one current and one state-dependent observable, we may use \eqref{fri-corr} to write the bound explicitly
\begin{align}
\frac{\av{J}^2}{\text{Var}_J \big(1- {\chi_{J,Z}}^2 \big)} \leq \frac{1}{2} \Delta S^\text{irr} \label{TUR-corr} ,
\end{align}
which is equivalent to the CTUR \eqref{correlation-TUR}.
This is very appealing from an experimental point of view:
Currents as in \eqref{current} depend on the velocity or transitions between the internal states. 
Since observing these requires a high time-resolution, such quantities are generally challenging to measure accurately.
The only exception are specific choices of the weighting functions, for which the time-integrated observable can be measured directly, for example the displacement of a particle.
By contrast, observables of the type \eqref{nc-observable}, which depend only on the position and the internal state can easily be evaluated from trajectory data.
The CTUR \eqref{TUR-corr} implies that, provided at least one current can be obtained from the measurement, we may use other, non-current observables to improve the lower bound on the entropy production.

\section{Optimal observables and stochastic entropy production} \label{sec-optimal-obs}
Given that the choice of the observable $Z$ in \eqref{TUR-corr} has a lot of freedom, a natural question is whether there exists an optimal observable which maximizes the bound.
This is equivalent to finding $Z$ such that magnitude of the Pearson coefficient $\chi(J,Z)$ becomes maximal for given $J$.
Unfortunately, we have not been able to solve this optimization problem in general.
However, there is one particular case, where we can obtain the solution  explicitly.
For a pure Langevin dynamics without internal states, we may consider the stochastic entropy production $\Sigma$ as the observable $J$.
This corresponds to the weighting function
\begin{align}
\bm{w}(\bm{x}) &= \bm{B}^{-1} \bm{\nu}^\text{st}(\bm{x}) \label{meanvel} .
\end{align}
As we show in Appendix \ref{app-optimal-obs}, in this case, the optimal choice for $Z$ is
\begin{align}
Z = \bar{\Sigma} = \int_0^\tau dt \ \bm{\nu}^\text{st,T}(\bm{x}(t)) \bm{B}^{-1} \bm{\nu}^\text{st}(\bm{x}(t)) \label{optimal-obs-entropy} .
\end{align}
This quantity can be interpreted as a local mean entropy production, i.~e.~the expected entropy production rate at position $\bm{x}(t)$ integrated along the trajectory.
Note that both $\Sigma$ and $\bar{\Sigma}$ have the entropy production $\Delta S_\text{irr}$ as their average value.
As it turns out, this choice turns \eqref{TUR-corr} into an equality,
\begin{align}
2 \Delta S_\text{irr} = \text{Var}_\Sigma \big( 1 - {\chi_{\Sigma,\bar{\Sigma}}}^2 \big) ,
\end{align}
which shows that this really is the optimal choice of $Z$.
We remark that this equality is equivalent to the equality $2 \Delta S_\text{irr} = \text{Var}(\delta \Sigma)$ with $\delta \Sigma = \Sigma - \bar{\Sigma}$ derived in Ref.~\cite{Dec20b}.
For general currents, while the optimal $Z$ could not be obtained explicitly, we note that the average current is expressed in terms of the local mean velocity as
\begin{align}
\av{J} = \tau \int d\bm{x} \ \bm{w}^\text{T}(\bm{x}) \bm{\nu}^\text{st}(\bm{x}) p^\text{st}(\bm{x}) .
\end{align}
Comparing this to \eqref{optimal-obs-entropy}, this suggests that a good choice for $Q$ may be
\begin{align}
Z = \bar{J} = \int_0^\tau dt \ \bm{w}^\text{T}(\bm{x}(t)) \bm{\nu}^\text{st}(\bm{x}(t)) \label{obs-optimal}.
\end{align}
This choice is the local mean value of the current, which has the same average as the current itself.

Further insight into the meaning of the optimal observable $Z$ can be gained from the following consideration.
Since the average of the observables $J$ and $\tilde{J} = J - Z$ exhibit the same scaling under continuous time-reversal
\begin{align}
\partial_\theta \av{J}^\theta = \partial_\theta \av{\tilde{J}}^\theta = \av{J},
\end{align}
they both satisfy a TUR
\begin{align}
\frac{\av{J}^2}{\text{Var}_J} \leq \frac{1}{2} \Delta S^\text{irr} \quad \text{and} \quad \frac{\av{J}^2}{\text{Var}_{\tilde{J}}} \leq \frac{1}{2} \Delta S^\text{irr} \label{TUR-2} .
\end{align}
Since the choice of $Z$ is arbitrary within the class of observables \eqref{nc-observable}, we may minimize the variance of $\tilde{J}$ with respect to $Z$,
\begin{align}
\frac{\av{J}^2}{\inf_Z \big(\text{Var}_{\tilde{J}}\big)} \leq \frac{1}{2} \Delta S^\text{irr}.
\end{align}
We may generalize this slightly by choosing $\tilde{J} = J - \alpha Z$, where $\alpha$ is a constant.
In this case, the minimization with respect to $\alpha$ can be done explicitly and yields
\begin{align}
\inf_\alpha \big(\text{Var}_{\tilde{J}}\big) = \text{Var}_J \big( 1 - {\chi_{J,Z}}^2 \big),
\end{align}
from which we readily obtain \eqref{TUR-corr} and finding the optimal observable corresponds to maximizing the Pearson coefficient.
Intuitively, the optimal observable is the state-dependent observable whose fluctuations most closely mimic those of the current $J$, thus minimizing the variance of $J-Z$.
The tightness of the TUR is thus limited by how closely the current $J$ can be emulated by a state-dependent observable $Z$, i.~e., the magnitude of the predictable part of $J$.
An extreme case are the fluctuations of the stochastic entropy around its local mean value $\delta \Sigma$.
This quantity is equivalent to the entropy production measured in terms of the stochastic time coordinate introduced in Ref.~\cite{Pig17}.
Its statistics are described by simple Brownian motion and thus its predictable part is zero and the corresponding TUR is an equality \cite{Dec20b}.

In the case of a pure Markov jump dynamics, the right-hand side of the \eqref{TUR-corr} can be replaced by a tighter bound
\begin{align}
\frac{\av{J}^2}{\text{Var}_J \big(1- {\chi_{J,Z}}^2 \big)} \leq \frac{1}{2} R \leq \frac{1}{2} \Delta S^\text{irr} \label{TUR-corr-markov}
\end{align}
with the quantity $R$ given by
\begin{align}
R = \tau \sum_{i,j} \frac{\big(W_{ij} p^\text{st}_j - W_{ji} p^\text{st}_i \big)^2}{W_{ij} p^\text{st}_j + W_{ji} p^\text{st}_i} \label{pseudo-entropy} .
\end{align}
The inequality $\Psi \leq \Delta S^\text{irr}$ is a straightforward consequence of the elementary inequality
\begin{align}
\frac{(a-b)^2}{a+b} \leq \frac{1}{2} (a-b) \ln \bigg(\frac{a}{b} \bigg) \label{log-mean},
\end{align}
which holds of arbitrary $a,b > 0$.
We note that the quantity $R$ has been introduced several times in the recent literature \cite{Dec18c,Ots20,Shi21} and been termed pseudo entropy production in Ref.~\cite{Shi21}.
Like the entropy production \eqref{entropy}, this quantity measures the degree to which the system is out of equilibrium.
As shown in Appendix \ref{app-optimal-obs}, the choice
\begin{align}
\omega_{ij} = 2 \frac{W_{ij} p^\text{st}_j - W_{ji} p^\text{st}_i}{W_{ij} p^\text{st}_j + W_{ji} p^\text{st}_i}
\end{align}
in the current \eqref{current} allows $J = \mathcal{R}$ to be interpreted as a stochastic version of the pseudo entropy in the sense that $\av{\mathcal{R}} = R$.
Further, choosing
\begin{align}
Z = \bar{\mathcal{R}} = \int_0^\tau dt \ \sum_i \frac{\Big(W_{i j(t)} - W_{j(t) i} \frac{p_i^\text{st}}{p_{j(t)}^\text{st}} \Big)^2}{W_{i j(t)} + W_{j(t) i} \frac{p_i^\text{st}}{p_{j(t)}^\text{st}}} \label{optimal-obs-entropy-markov},
\end{align}
that is, the local mean of the current $\mathcal{R}$, we obtain equality in \eqref{TUR-corr-markov},
\begin{align}
2 R = \text{Var}_\mathcal{R} \big(1- {\chi_{\mathcal{R},\bar{\mathcal{R}}}}^2 \big) ,
\end{align}
which is the analog of \eqref{optimal-obs-entropy}.
This shows that as in the Langevin case, we may realize equality in \eqref{TUR-corr-markov}, with the difference that the quantity being estimated is the pseudo entropy production $R$ instead of the entropy production $\Delta S^\text{irr}$.
As a consequence, \eqref{TUR-corr-markov} can only yield a reasonable estimate on the entropy production if $\Delta S^\text{irr} \approx R$.
From \eqref{log-mean}, this holds when $|W_{ij} p^\text{st}_j - W_{ji} p^\text{st}_i| \ll W_{ij} p^\text{st}_j + W_{ji} p^\text{st}_i$ for all transitions, i.~e.~the bias across any transition is small compared to the total activity.
The latter condition is realized either near equilibrium (where the total bias is small) or in the continuum limit (where a finite total bias is distributed over many individual transitions).
In all other cases, \eqref{TUR-corr-markov} can at most yield a lower bound on $\Delta S^\text{irr}$, the extreme case being unidirectional transitions \cite{Pal21}, where $\Delta S^\text{irr}$ diverges while $R$ remains finite.
Note that, comparing \eqref{optimal-obs-entropy-markov} with the definition of a general current \eqref{current} suggests that a good choice for the state-dependent observable should be
\begin{align}
Z = \bar{J} = \int_0^\tau dt \ \sum_{i} \omega_{i j(t)} V^\text{st}_{i j(t)}, \label{obs-optimal-markov}
\end{align}
with the \enquote{local mean velocity}
\begin{align}
V_{ij}^\text{st} = \frac{1}{2} \bigg( W_{i j} - W_{j i} \frac{p^\text{st}_i}{p^\text{st}_j} \bigg) .
\end{align}

\section{Improved TUR from trajectory data} \label{sec-data}
Based on the considerations in the previous sections, we now provide a general recipe to obtain sharper TUR-like bounds from existing trajectory data.
The goal of this procedure is to use the trajectory data to obtain an estimate of the entropy production.
We assume that the current $J$, i.~e.~the weighting functions $\bm{w}_i(\bm{x})$ and $\omega_{i j}(\bm{x})$ in \eqref{current} are fixed.
The reasoning behind this is that, in order to optimize the current, we need to explicitly observe all the transitions along the trajectory, which typically requires a sufficiently high time-resolution.
Failing that, we are limited to observing current observables whose time-integral directly corresponds to a measurable quantity, for example the total displacement of a tracer particle.
Then, obtaining a good estimate for the entropy production corresponds to finding a good choice for the observable $Z$ in \eqref{TUR-corr}.
We propose three methods for finding such a choice.
We also point the reader to Appendix \ref{app-error}, where we investigate the dependence of the estimate on the size of the data set and the sampling interval.

\subsection{Explicit optimization} \label{sec-data-numopt}
Given a set of trajectory data, we may in principle optimize the function $z_i(\bm{x})$ in \eqref{nc-observable} explicitly by maximizing the Pearson coefficient between $J$ and $Z$.
In the case of a pure jump process with $M$ states, this corresponds to optimizing $M-1$ parameters---since the Pearson coefficient is invariant under a global rescaling of $Z$, we may set $z_1 = 1$ without loss of generality.
In the case of a diffusion process, we may choose a reasonable set of (say $K$) basis functions $f_k(\bm{x})$ and write $z(\bm{x}) = \sum_{k=1}^K c_k f_k(\bm{x})$, which (setting $c_1 = 1$) again corresponds to optimizing $K-1$ parameters.
Provided that $K$ and the data set are not too large, this explicit optimization is feasible and has the advantage of yielding the tightest possible bound within the accuracy of the optimization algorithm.
However, as $K$ or the size of the data set increase, this procedure becomes increasingly unfeasible and less tight but easier to compute heuristic bounds may be desired.

\subsection{Approximate local mean velocity} \label{sec-data-meanvel}
Given that, in many physically relevant situations, the local mean current \eqref{obs-optimal} is expected to give a reasonably tight bound, we may use an approximate expression $\tilde{\bm{\nu}}^\text{st}(\bm{x})$ for the local mean velocity to compute $Z$.
Such an approximate expression may be obtained directly from the trajectory data.
Even if the measurement resolution is not sufficient to observe every transition, we may still obtain a coarse-grained mean velocity profile.
For a spatial resolution $\Delta x$ and a temporal resolution $\Delta t$, we may divide the observation volume into cells of size $\Delta x^N$ and obtain an estimate for the local mean velocity in a given cell by considering the displacement at time $t+\Delta t$, averaged over all particles starting out in the cell at time $t$.
An approximate local mean velocity may also be obtained from theoretical considerations by considering a simplified version of the dynamics, in which the local mean velocity can be computed explicitly and is expected to resemble the one of the actual dynamics.
A simplified model may be obtained, for example, by linearizing a given non-linear model or by constructing a lower-dimensional effective model.
Even though both approaches are generally valid only in a certain limit, in many cases, the resulting local mean velocity still yields an improved bound via the CTUR even beyond the strict validity of the approximations.
As a concrete example, we consider in the next section the flow through a two-dimensional channel, which we approximate by a one-dimensional flow in an effective potential, for which an explicit expression of the local mean velocity can be obtained.

Another application, for which we anticipate this approach to be useful, is for systems in the presence of spatial disorder.
Since the disorder configuration is generally not known precisely, we also cannot obtain an explicit expression for the local mean velocity.
In this case, we may compute the local mean velocity for the disorder-free system, and provided that the disorder is not too strong, we can still expect the CTUR with the corresponding observable $Z$ to provide a notable improvement over the TUR.
For example, we can imagine a channel as in Fig.~\ref{fig-channel-potential}, through which a current is driven by an external bias.
However, due to variances in the fabrication of the channel or the presence of contamination, the actual potential experienced by the particle may be more complicated.
Nevertheless, if we can obtain an expression for the local mean velocity of the ideal situation, we can use it to obtain an improved estimate of the entropy production.

\subsection{Inverse occupation fraction} \label{sec-data-occ}
Finally, we provide a concrete formula for obtaining a candidate for the observable $Z$.
This relies on the observation that, in many systems, the local mean velocity at some position $\bm{x}$ and the probability of finding the particle at position $\bm{x}$ are approximately inversely proportional.
This relation is exact for Langevin dynamics in a spatially periodic one-dimensional system; in this case, we have $\nu^\text{st}(x) p^\text{st}(x) = v^\text{d}/L$, where $v^\text{d}$ is the drift velocity and $L$ the period of the system.
The same relation holds for a Markov jump dynamics with on a ring, $V^\text{st}_{i \pm 1,i} p_i^\text{st} = \pm \mathcal{J}/N$ where $\mathcal{J}$ is the total current on the ring.

However, even in higher-dimensional settings, we often expect the probability density $p^\text{st}(\bm{x})$ to be small where the magnitude of $\bm{\nu}^\text{st}(\bm{x})$ is large and vice versa.
The reasoning for this relation is that, for ergodic dynamics, $p^\text{st}(\bm{x}) d\bm{x}$ is equal to the temporal occupation fraction, i.~e.~the fraction of time that the particle resides in the volume element $d\bm{x}$ around $\bm{x}$ in the limit of long observation times.
On the other hand, if the local mean velocity at $\bm{x}$ is large, this means that the particle tends to move at a large velocity at $\bm{x}$ and thus we expect it to only spend a short time in the volume element $d\bm{x}$.
Note that this qualitative argument may break down, e.~g.~near vortices in the local mean velocity, where particles may move fast yet still mostly reside in a small volume.
However, since we only require an approximation to the local mean velocity to improve the TUR bound on the entropy production, using the time-integral of the inverse occupation fraction as the observable $Z$ is still expected to yield a notable improvement in most cases.
Specifically, dividing the observation volume into $\mathcal{N}$ cells (for simplicity, we assume they have equal volume $v$) and the observation time into $\mathcal{M}$ steps, we define as the occupation number $\mathcal{K}_i$ the number of times that any trajectory takes a value in the $i$-th cell.
Then, we may define the observable $Z$ as
\begin{align}
Z = \sum_{m = 1}^\mathcal{M} \frac{1}{\mathcal{K}_{i(m)}} \label{occupation-observable},
\end{align}
where $i(m)$ is the cell the trajectory is located in at time step $m$.
Note that, while $Z$ is evaluated for each individual trajectory, $\mathcal{K}_i$ is computed from the entire set of trajectory data.
We do not need to normalize the occupation number into an occupation fraction, since the overall normalization factor does not impact the Pearson coefficient.
While the observable \eqref{occupation-observable} may not yield the tightest bound in \eqref{TUR-corr}, it has the advantage that it may be computed using only the trajectory data without any additional input.

Finally, we remark upon a possible issue with the definition \eqref{occupation-observable}.
If the phase-space volume $\mathcal{V}$ that is accessible to the system is not finite, then it becomes necessary to introduce a lower cutoff on the occupation number $\mathcal{K}_i$.
To see this, note that for an ergodic system in the limit of long time and/or a large number of trajectories, we have $\mathcal{K}_i \propto p^\text{st}(\bm{x}_i) v$, where $\bm{x}_i$ is the location of the $i$-th cell and $v$ is the cell volume.
Then,
\begin{align}
\av{Z} \propto \int_{\mathcal{V}} d\bm{x} \ \frac{1}{p^\text{st}(\bm{x})} p^\text{st}(\bm{x}) = \mathcal{V} .
\end{align}
So the average of $Z$ is proportional to the volume $\mathcal{V}$ and diverges if the latter becomes infinite.
For example, if $p^\text{st}(\bm{x}) \propto \exp(-\Vert \bm{x} \Vert^2/(2 r_0))$, then $1/p^\text{st}(\bm{x})$ grows exponentially for large $\Vert \bm{x} \Vert$ and its average is not well-defined.
Intuitively, this means that \eqref{occupation-observable} is dominated by rare events, which are never well sampled no matter how many trajectories we record.
To avoid this problem, we may introduce a lower cutoff on $\mathcal{K}_i$, that is, we only count cells that have been visited in at least a fraction $\epsilon$ of all data points.
In terms of the probability density, this means that we discard all points with $p^\text{st}(\bm{x}) < \frac{\epsilon}{v}$.
In doing so, the average of $Z$ becomes finite and we may use the resulting observable as a qualitative estimate for the local mean velocity.
Obviously, the resulting quantity is only an approximation to the actual occupation fraction, however, we may still expect it to yield a useful improvement over the TUR, provided that $\epsilon$ is sufficiently small and the statistics of the current are not dominated by rare events.
This lower cutoff also avoids another possible issue: 
In \eqref{occupation-observable}, we use the same data set to evaluate the occupation fraction and the trajectory-dependent observable $Z$.
While this is not problematic for a large number of trajectories, since the influence of any single trajectory on $\mathcal{K}_i$ is small, it may lead to unintended correlations if the number of trajectories is not large.
In the latter case, it may be preferable to divide the data set into two parts, using one to determine $\mathcal{K}_i$ and the other to compute $Z$.
In doing so, we may encounter the situation where a given cell is visited in the second part but not the first part of the data set.
In this case, \eqref{occupation-observable} is no-longer well defined since we divide by zero. 
Introducing a lower cutoff excludes this possibility.
However, we stress that even if $\mathcal{K}_i$ and $Z$ are computed from the same data set, \eqref{correlation-TUR} remains valid, so the worst outcome of unintended correlations is to reduce the quality of the estimate on the entropy production.

\section{Demonstrations} \label{sec-demonstration}

\subsection{Molecular motor model} \label{sec-demonstration-motor}
To demonstrate how \eqref{TUR-corr} may be used to obtain a tight bound on the entropy production, we consider the model for the $\text{F}_1$-ATPase molecular motor introduced in Ref.~\cite{Zim12} and further studied in Ref.~\cite{Kaw14}.
This model describes the motion of a probe bead coupled to a rotating molecular motor, which either consumes a molecule called ATP in order to generate rotational motion, or, conversely, generates ATP when driven by an external torque.
The probe is considered to be trapped inside a potential $U_i(x)$, which is determined both the joint between the probe and the motor and the internal structure of the motor.
As the motor rotates in steps of length $L$, the potential depends on the current state of the motor as $U_i(x) = U_0(x - i L)$.
In the simplest form of the model, the trapping potential is harmonic, $U_0(x) = k x^2/2$, and the motor rotates in steps of $L = 120^\circ$.
The transitions between the states of the motor are described by the position-dependent transition rates
\begin{align}
W_{i}^+ &= W_0 \text{exp}\Big[\frac{\alpha}{\kb T} \big( U_i(x) - U_{i+1}(x) + \Delta \mu \big) \Big], \\
W_{i+1}^- &= W_0 \text{exp}\Big[\frac{1-\alpha}{\kb T} \big( U_{i}(x) - U_{i+1}(x) + \Delta \mu \big) \Big] \n ,
\end{align}
where $W_i^+$ ($W_{i+1}^-$) is the rate of transitions from $i$ to $i+1$ (from $i+1$ to $i$). 
Here, $W_0$ quantifies the overall activity of the motor, which is proportional to the concentration of ATP in the environment of the motor.
$\Delta \mu$ is the chemical potential difference driving the rotation; it is the amount of energy gained by the motor when consuming a single molecule of ATP.
The parameter $0 \leq \alpha \leq 1$ characterizes the asymmetry of the position-dependence of the rates.
The spatial motion of the probe is described by one-dimensional Brownian motion in the potential $U_i(x)$.
In addition, an external torque may be applied to the probe, which enters the equation of motion in the form of a non-conservative bias force $F$.
As a consequence, the drift coefficient in \eqref{langevin} is given by $a_i(x) = (-U'_i(x) - F)/\gamma$, where $\gamma$ is the friction coefficient, $F$ is an external force acting on the probe, while the diffusion matrix is $G = \sqrt{2 \kb T/\gamma}$.
An experimentally accessible current is the total displacement of the probe,
\begin{align}
J = \int_0^\tau dt \ \dot{x}(t) .
\end{align}
In the steady state, we have $J = v^\text{d} \tau$, where $v^\text{d}$ is the drift velocity.
Because the system is effectively one-dimensional, the local mean velocity in the steady state is given by $\nu^\text{st}(x) p^\text{st}(x) = v^\text{d}/L$ where $p^\text{st}(x)$ is the $L$-periodic steady state probability density.
In this case, we can thus reconstruct the local mean velocity from the trajectory data of the probe by taking a histogram of the probe positions.
However, it should be noted that this local mean velocity is a coarse-grained quantity, the true local mean velocity entering \eqref{fpe} also depends on the state $i$ of the motor.
We define the observable
\begin{align}
Z = \int_0^\tau dt \ \frac{1}{p^\text{st}(x(t))} ,
\end{align}
which is proportional to $\bar{J}$.
Since the proportionality factor cancels in the Pearson coefficient, $Z$ and $\bar{J}$ are equivalent with respect to \eqref{TUR-corr}.
To asses the tightness of the various inequalities, we introduce the transport efficiencies
\begin{align}
\eta_J &= \frac{2 \av{J}^2}{\text{Var}_J \Delta S^\text{irr}} , \label{transport-eff} \\
\eta_{J,\bar{J}} &= \frac{2 \av{J}^2}{\text{Var}_J \Delta S^\text{irr} \big(1 - {\chi_{J,\bar{J}}}^2 \big)} \n .
\end{align}
Both of these quantities are smaller than unity and measure the magnitude of the average transport relative to its fluctuations and the dissipation.
\begin{figure*}
\includegraphics[width=.49\textwidth]{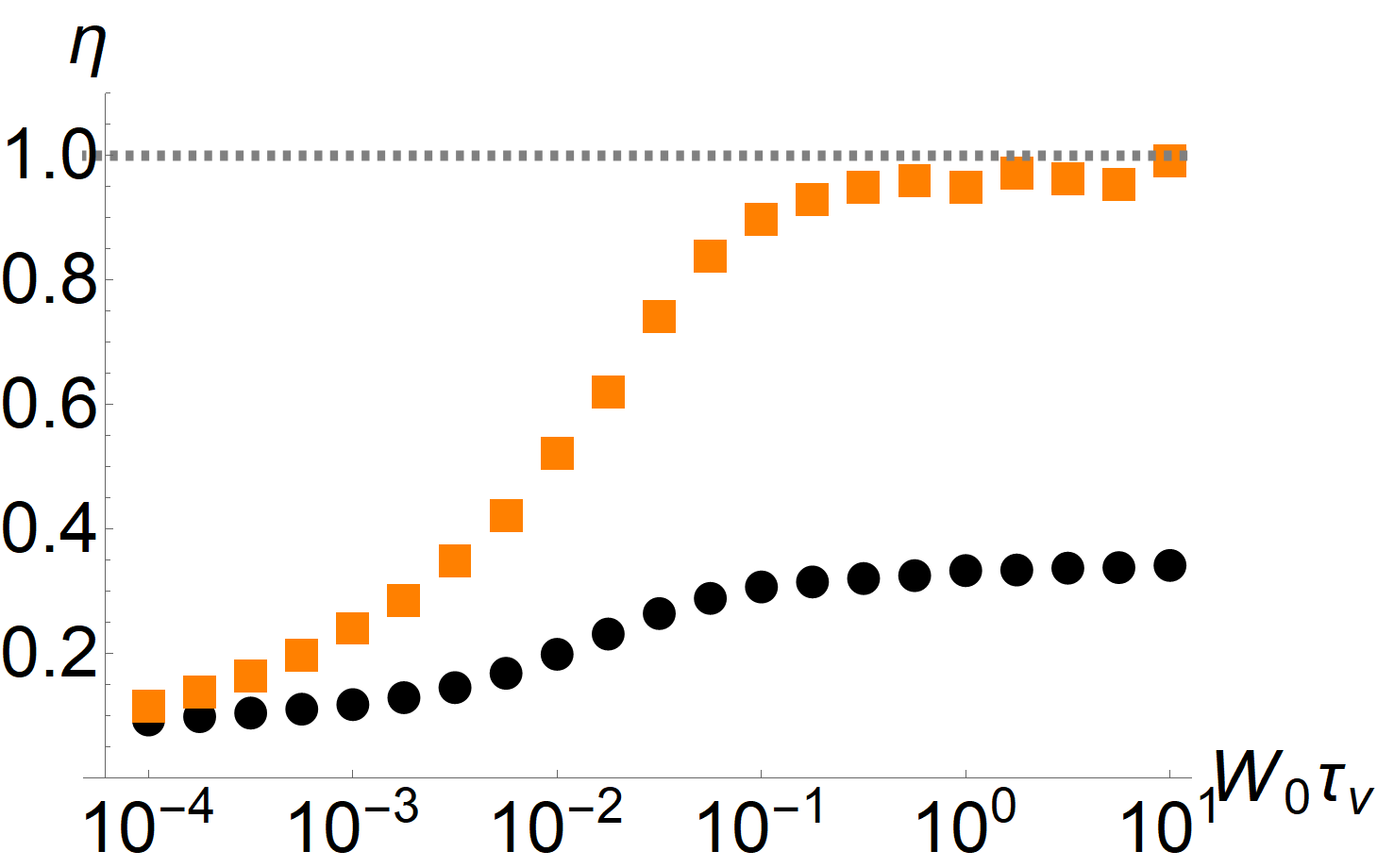}
\includegraphics[width=.49\textwidth]{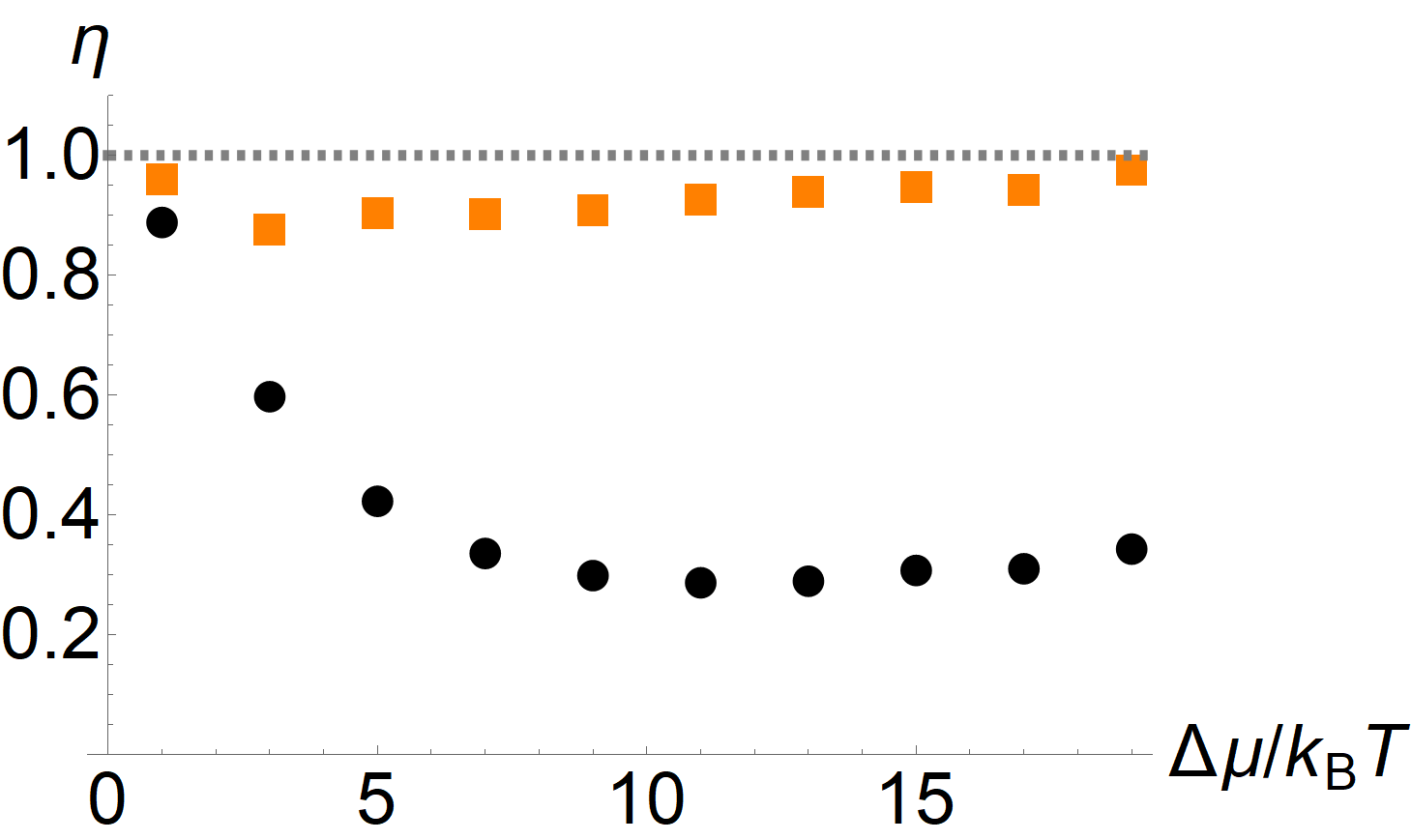}\\
\includegraphics[width=.49\textwidth]{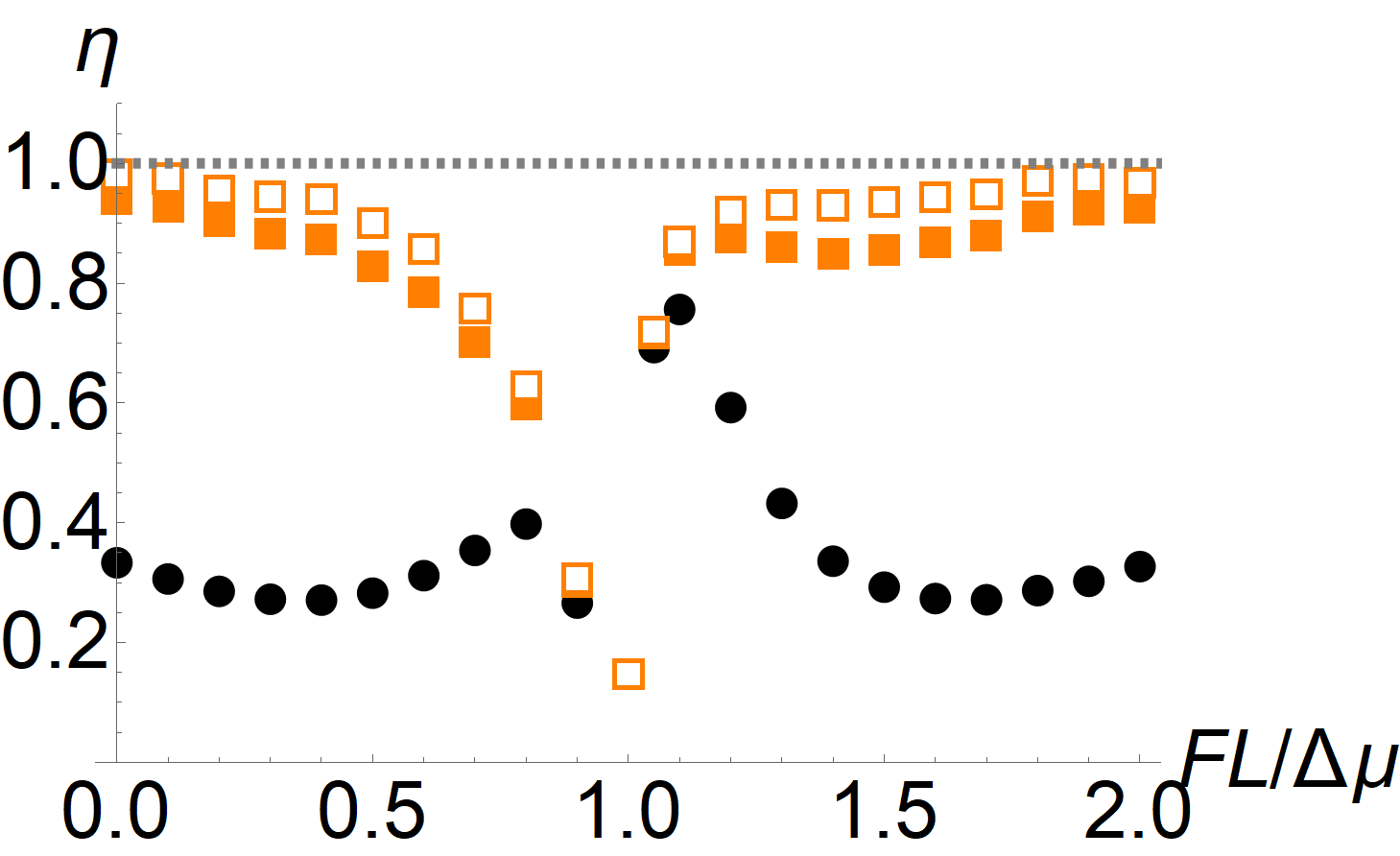}
\includegraphics[width=.49\textwidth]{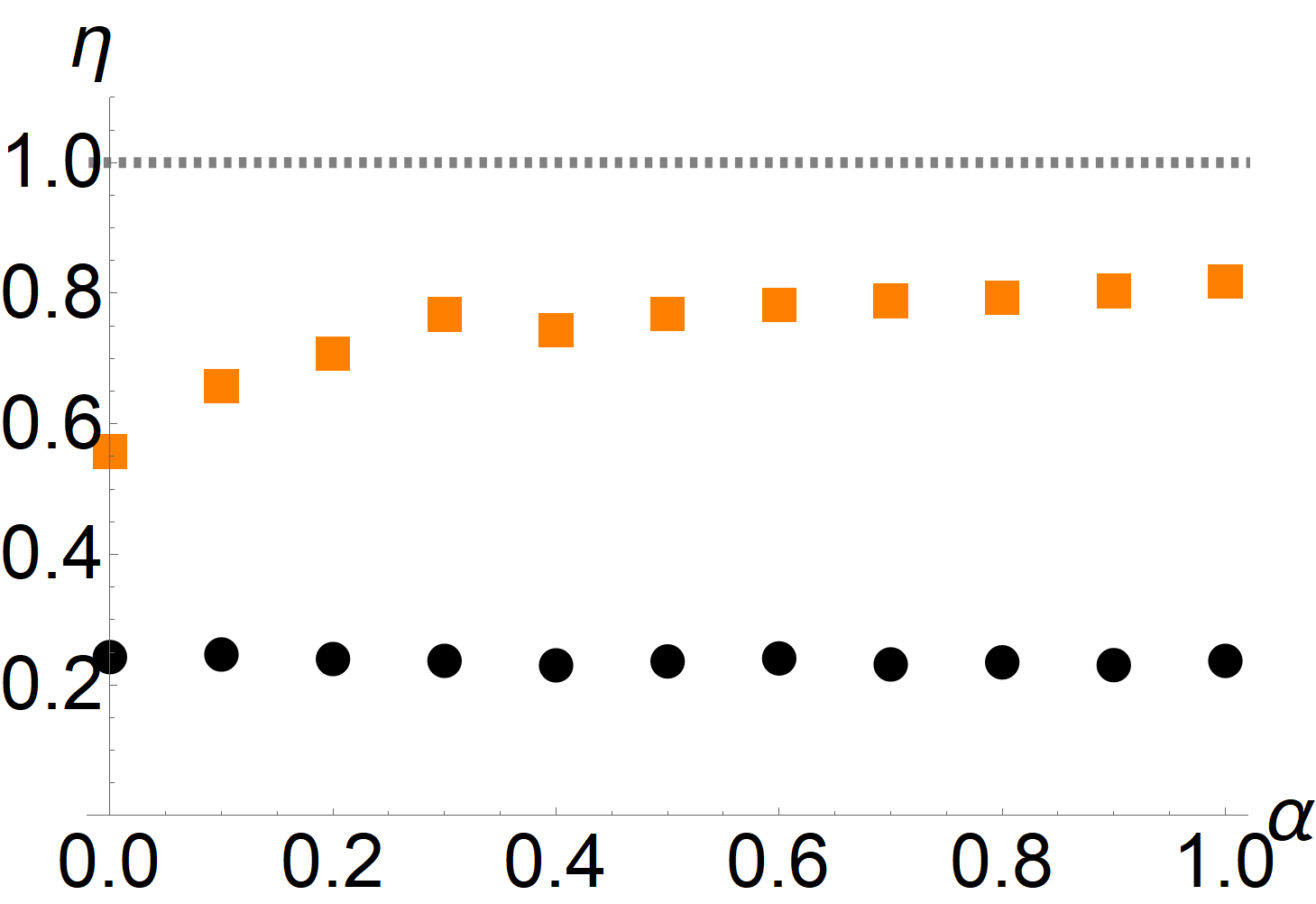}
\caption{The transport efficiency \eqref{transport-eff} for the displacement of the probe attached to the molecular motor as a function of different parameters.
The black circles correspond to the TUR for the displacement $z$ only, while the solid orange squares show the CTUR \eqref{TUR-corr} including the correlations between the displacement and its local mean value $\bar{z}$. 
The data are obtained by numerical simulations of \eqref{langevin} with $W_0 \tau_\text{v} = 10$, $U_0 = 50 k_B T$, $\Delta \mu = 19 k_B T$, $\alpha = 0.1$, $\gamma = 2.5 \cdot 10^3$ and $F = 0$, except where noted differently.
(Top left) As a function of the base activity $W_0$. The horizontal axis is scaled by the timescale $\tau_v = \gamma \kb T/(k L)^2$ \cite{Kaw14}.
(Top right) As a function of the chemical potential difference $\Delta \mu$.
(Bottom left) As a function of the external load $F$. The empty orange squares are \eqref{TUR-corr} with a numerically optimized observable, see \eqref{parameter-observable}.
(Bottom right) As a function of the asymmetry parameter $\alpha$.  \label{fig-motor-freq}}
\end{figure*}
The efficiencies \eqref{transport-eff} are shown for the molecular motor model in Fig.~\ref{fig-motor-freq} as a function of various parameters.
The top-left panel shows $\eta$ as a function of the base activity $W_0$, which corresponds to the concentration of ATP in the experiment.
For small activity both $\eta_J$ and $\eta_{J,\bar{J}}$ are comparable and small; in this limit, the transitions between the different motor conformations are not translated efficiently into motion and the dissipation is not reflected in the motion of the probe \cite{Kaw14}.
For large activity, $\eta_J$ saturates at a value of around $0.4$.
However, when we compute $\eta_{J,\bar{J}}$ in this regime, we find that it saturates at a value close to unity, i.~e.~the maximum possible value.
The top-right panel shows $\eta$ as a function of the chemical potential difference $\Delta \mu$.
While this value cannot be readily changed in experiment, it yields important insight into the nature of the bound \eqref{TUR-corr}.
For small $\Delta \mu$ the system is almost in equilibrium, and both the TUR and CTUR are close to an equality.
However, as we drive the system out of equilibrium, the TUR ratio quickly drops, while the CTUR remains close to unity.
This suggests that, while the TUR is generically only saturated close to equilibrium \cite{Mac18,Li19}, the improved bound from the CTUR \eqref{TUR-corr} can yield an accurate estimate of the entropy production even far from equilibrium.
The bottom-left panel shows $\eta$ as a function of the external load force.
Close to the stall condition $F L = \Delta \mu$ neither of the bounds is tight.
This is reasonable, since when the motor stalls, also the probe stops moving, while the motor keeps changing its conformation and thus dissipating energy.
Interestingly, both bounds are close to unity slightly above the stall condition, i.~e.~when the external load is just strong enough to turn the motor in the opposite direction.
Away from the stall condition, we again find that the TUR is rather loose, while the CTUR remains tight.
Note that in this panel we also included the results obtained by optimizing the observable $Z$.
Specifically, we write 
\begin{align}
\tilde{Z} = \int_0^\tau dt \sum_{k=1}^K \bigg( a_k \sin\Big(\frac{2 \pi x(t)}{L} \Big) + b_k \cos\Big(\frac{2 \pi x(t)}{L} \Big) \bigg) \label{parameter-observable}
\end{align}
and then numerically optimize the parameters $a_k$, $b_k$ such that $\chi(J,\tilde{Z})^2$ is maximal for the given trajectory data.
Here we use $K = 10$; further increasing of the number of parameters provides no notable improvement. 
The numerical optimization is done using Mathematica's \texttt{NMaximize} command.
The result are the empty orange squares in the bottom-left panel of Fig.~\ref{fig-motor-freq}.
As can be seen, the observable $\bar{Z}$ is not truly optimal, so that some improvement of the lower bound on entropy production is possible.
However, the heuristic choice $\bar{Z}$ already provides a useful estimate without the need for any parameter optimization.
Finally, the bottom-right panel shows $\eta$ as a function of the asymmetry parameter $\alpha$.
As was shown in Ref.~\cite{Kaw14} the motor can operate without internal dissipation close to $\alpha = 0$, whereas other values of $\alpha$ result in a finite amount of internal dissipation.
Since this difference is most pronounced at low activity, we choose $W_0$ such that the velocity remains constant at $v = 0.65 v_\text{max}$ for all values of $\alpha$; this corresponds to $W_0 \tau_v \approx 10^{-2}$ in the top-left panel.
While in this regime of low activity, neither bound is saturated, \eqref{TUR-corr} does yield a considerable improvement over the TUR.
Interestingly, neither bound shows a pronounced dependence on $\alpha$, which indicates that their tightness, at least for this model, is not related to the amount of internal vs. external dissipation.

The molecular motor model is essentially one-dimensional: Since the energy scale of the coupling between the probe particle and motor are large compared to the temperature, the position of the probe and the state of the motor are tightly coupled, so measuring the displacement of the probe is almost equivalent to measuring the internal transitions of the motor.
Indeed, considering the internal transitions of the motor as an additional current $J_2$, \eqref{TUR-multi} yields no notable improvement over the TUR.
Because of this, it is reasonable that the entropy production can be estimated accurately from a measurement of the probe position.
However, this tight coupling breaks down for slow switching (see the top-left panel of Fig.~\ref{fig-motor-freq}): In this regime, while the asymmetry in the transitions between the motor states still gives rise to entropy production, these transitions increasingly fail to produce a directed motion of the probe and thus its displacement can no longer be used to obtain an accurate estimate of entropy production.

In summary, we find that the TUR is generally not tight for this model, which mirrors the behavior in other types of molecular motors \cite{Hwa18}.
Viewed on its own, this would suggest that $\text{F}_1$-ATPase is not efficient at saturating the bound set by the TUR.
When we take into account the correlations between the velocity and its local mean value via the CTUR, the resulting inequality is almost saturated. 
Thus, in reality, the motor operates close to the limit permitted by the thermodynamic bound in the biologically relevant parameter regime.
In contrast to the TUR, which represents a trade-off between dissipation and precision, the biological meaning of saturating the CTUR is not so clear.
We note that, in the case of a particle moving in a flat potential, where the local mean velocity is independent of space, the TUR and CTUR are equivalent and both saturated.
The presence of a spatially varying local mean velocity generally increases the amount of fluctuations in the current and also induces correlations between the former and the latter.
As a consequence, the TUR becomes less tight and we explicitly need to take into account the correlations to counter this.
However, in case of the $\text{F}_1$-motor, the spatial inhomogeniety of the dynamics is also an integral part of the mechanism creating the motion of the motor in the first place.
In that sense, the CTUR may be interpreted as a trade-off between dissipation, precision and the spatial structure necessary for the motion of the motor.

\subsection{Two-cycle Markov jump process} \label{sec-demonstration-markov}
\begin{figure}
\includegraphics[width=.49\textwidth]{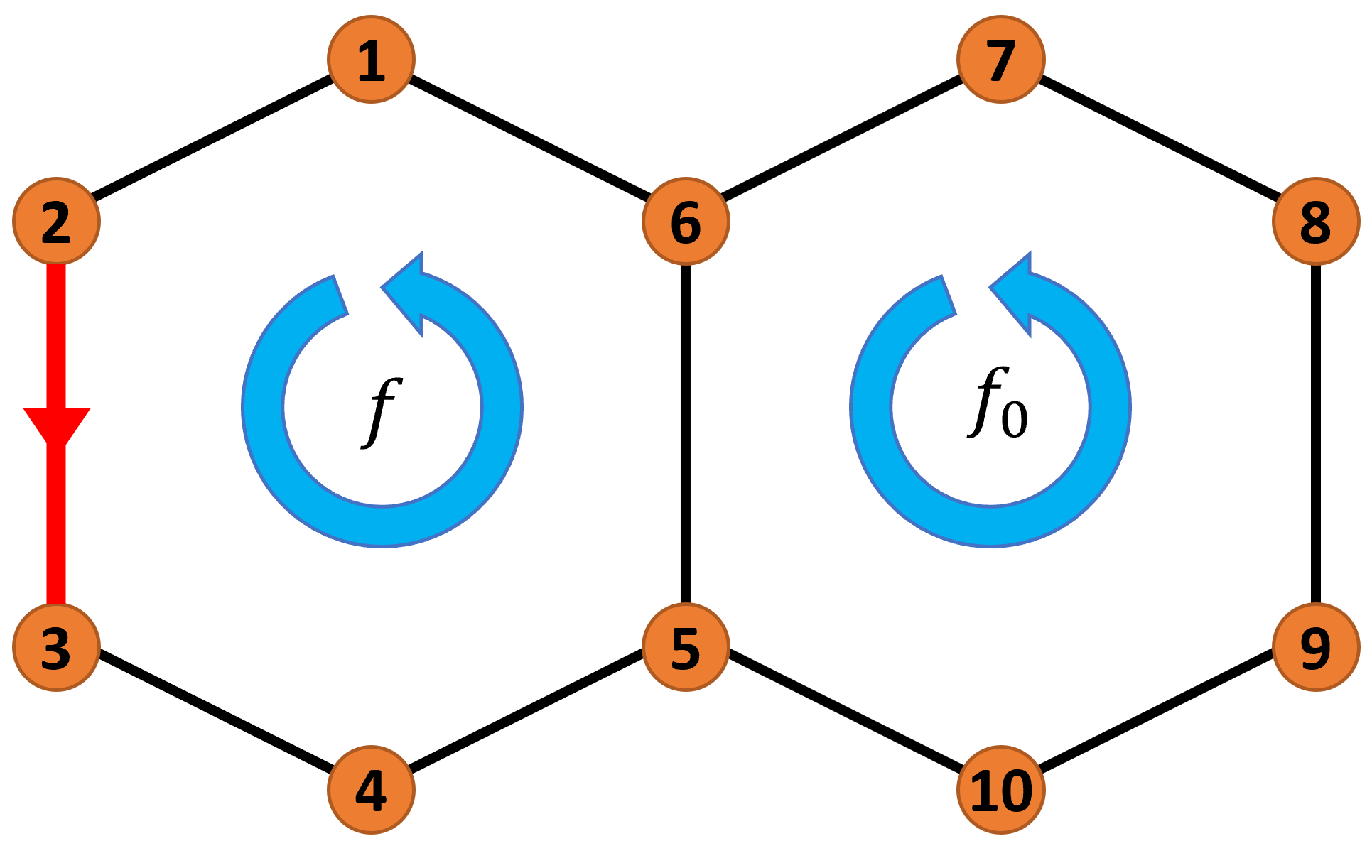}
\caption{An illustration of the two-cycle model Markov jump model. The model consists of $10$ arranged in two rings, with sites with sites $5$ and $6$ shared between the two rings. The transitions between the sites are given by the rates \eqref{two-cycle-rates} with a bias $f$ in the left and $f_0$ in the right ring. The current is measured between the sites $2$ and $3$ (red arrow).  \label{fig-two-cycle}}
\end{figure}

An obvious question is how the CTUR fares in the presence of multiple independent currents.
To investigate this issue, we introduce a simple Markov jump model consisting of $N=10$ sites forming two loops with a common edge, see Fig.~\ref{fig-two-cycle}.
While such two-cycle models are also used in modeling the chemical state space of molecular motors \cite{Lie07}, we use a simpler setup in order to better understand its behavior in relation to the CTUR.
We parameterize the transition rates between connected states as
\begin{align}
W_{ij} = \exp \bigg[ \frac{\beta}{2} \big( E_{i} - E_{j} \pm f \big) \bigg] \label{two-cycle-rates},
\end{align}
where $\beta$ is the inverse temperature, $E_i$ the energy of state $i$ and $f$ is the bias, with the choice $+f$ corresponding to counter-clockwise and $-f$ to clockwise transitions.
As indicated in Fig.~\ref{fig-two-cycle}, we keep the bias of the right cycle fixed at a value $f_0$ while varying the bias $f$ in the left cycle; the bias on the shared link $5 \rightarrow 6$ is set to $f - f_0$.
As the current $J_\text{L}$, we choose the number of counterclockwise transitions through the link $2 \rightarrow 3$ (indicated in red in Fig.~\ref{fig-two-cycle}), i.~e.~$\omega_{32} = 1 = -\omega_{23}$ and $\omega_{i j} = 0$ otherwise in \eqref{current}.
\begin{figure}
\includegraphics[width=.49\textwidth]{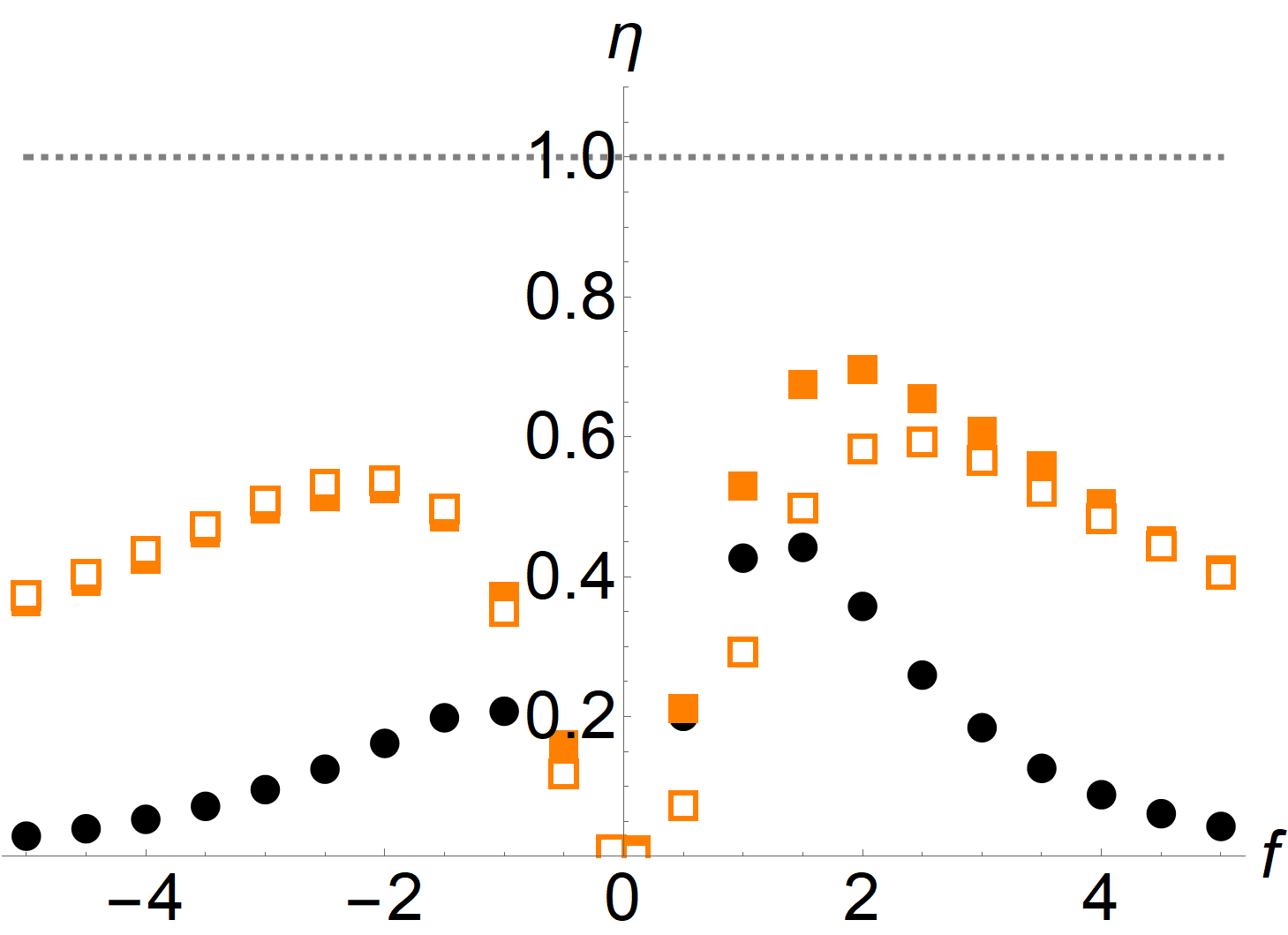}\\
\includegraphics[width=.49\textwidth]{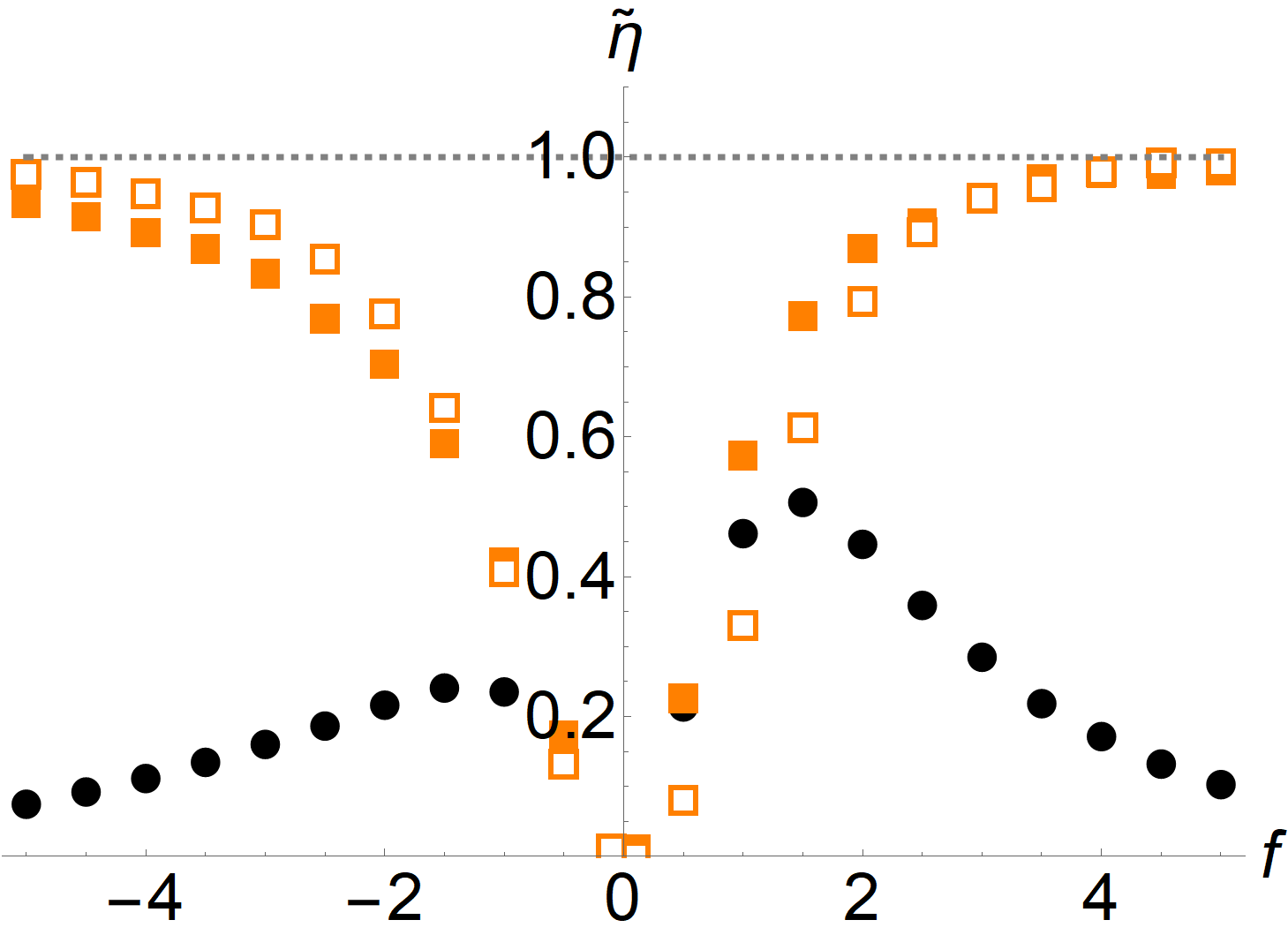}
\caption{The ratios $\eta$, \eqref{transport-eff-two-cycle}, and $\tilde{\eta}$, \eqref{transport-eff-two-cycle-pseudo}, for the two-cycle model as a function of the bias in the left cycle. The filled black circles correspond to the TUR bound $\eta_J$ and the filled orange squares to the CTUR bound $\eta_{J,Z}$, in both cases for a flat energy landscape $E_i \equiv 1$, $\beta = 1$ and $f_0 = 1$. The empty orange squares are the results for a random energy landscape, where $E_i$ for each site is randomly chosen between $0$ and $3$. All data are obtained from Monte-Carlo simulations using the rates \eqref{two-cycle-rates}. \label{fig-two-cycle-result}}
\end{figure}
As the state-dependent observable $Z_\text{L}$, we choose the inverse of the steady state occupation fraction in the left ring, $1/p^\text{st}_{i \vert L}$, which corresponds to measuring the ratio of the time spent in state $i$ and the time spent in the left ring.
The motivation behind this choice is that, even though the system no longer consists of a single cycle, we still expect the occupation fraction to be approximately inversely proportional to the current through the site, as argued in Section \ref{sec-data-occ}.
This observable can be measured by observing only the states in the left ring, for example if the right ring cannot be directly observed.
As in the previous example, we define the ratios
\begin{align}
\eta_J &= \frac{2\av{J}^2}{\text{Var}_J \Delta S^\text{irr}}, \label{transport-eff-two-cycle}\\
\eta_{J,Z} &= \frac{2\av{J}^2}{\text{Var}_J \Delta S^\text{irr} \big(1-\chi_{J,Z}^2 \big) } \n .
\end{align}
which are shown in Fig.~\ref{fig-two-cycle-result}.
In the absence of bias, $f = 0$, the average current $J_\text{L}$ in the left ring vanishes and thus neither the TUR nor the CTUR give us a non-trivial estimate on the entropy production.
The contribution of $J_L$ to the total entropy production increases with increasing bias $f$ and thus we can get a non-trivial lower bound on the entropy production.
Quantitatively, we observe that the estimate provided by the CTUR is considerably more accurate than the TUR, in particular for larger bias.
This finding is independent of the precise potential landscape; indeed, the bound obtained from the CTUR for flat and random configurations of energies $E_i$ is qualitatively similar.
However, as noted in \eqref{TUR-corr-markov}, in case of a Markov jump process, we are actually estimating the pseudo entropy production \eqref{pseudo-entropy} $R$.
Since the latter is smaller than the entropy production $\Delta S^\text{irr}$ for large bias (i.~e.~asymmetry between forward and reverse transitions), the estimate becomes worse when we increase the bias further, even though most of the entropy production in the system now stems from the left ring.
To quantify the difference between $R$ and $\Delta S^\text{irr}$, we also consider the modified efficiencies
\begin{align}
\tilde{\eta}_J &= \frac{2\av{J}^2}{\text{Var}_J  R}, \label{transport-eff-two-cycle-pseudo}\\
\tilde{\eta}_{J,Z} &= \frac{2\av{J}^2}{\text{Var}_J  R \big(1-\chi_{J,Z}^2 \big) } \n .
\end{align}
As can be seen from Fig.~\ref{fig-two-cycle-result}, the CTUR saturates this bound in the limit of large bias; thus, we can accurately estimate the pseudo entropy production from a measurement of the current and the occupation probabilities in the left ring.
By contrast, just as for the molecular motor model, the TUR estimate becomes less accurate in the limit of large bias.
\begin{figure}
\includegraphics[width=.49\textwidth]{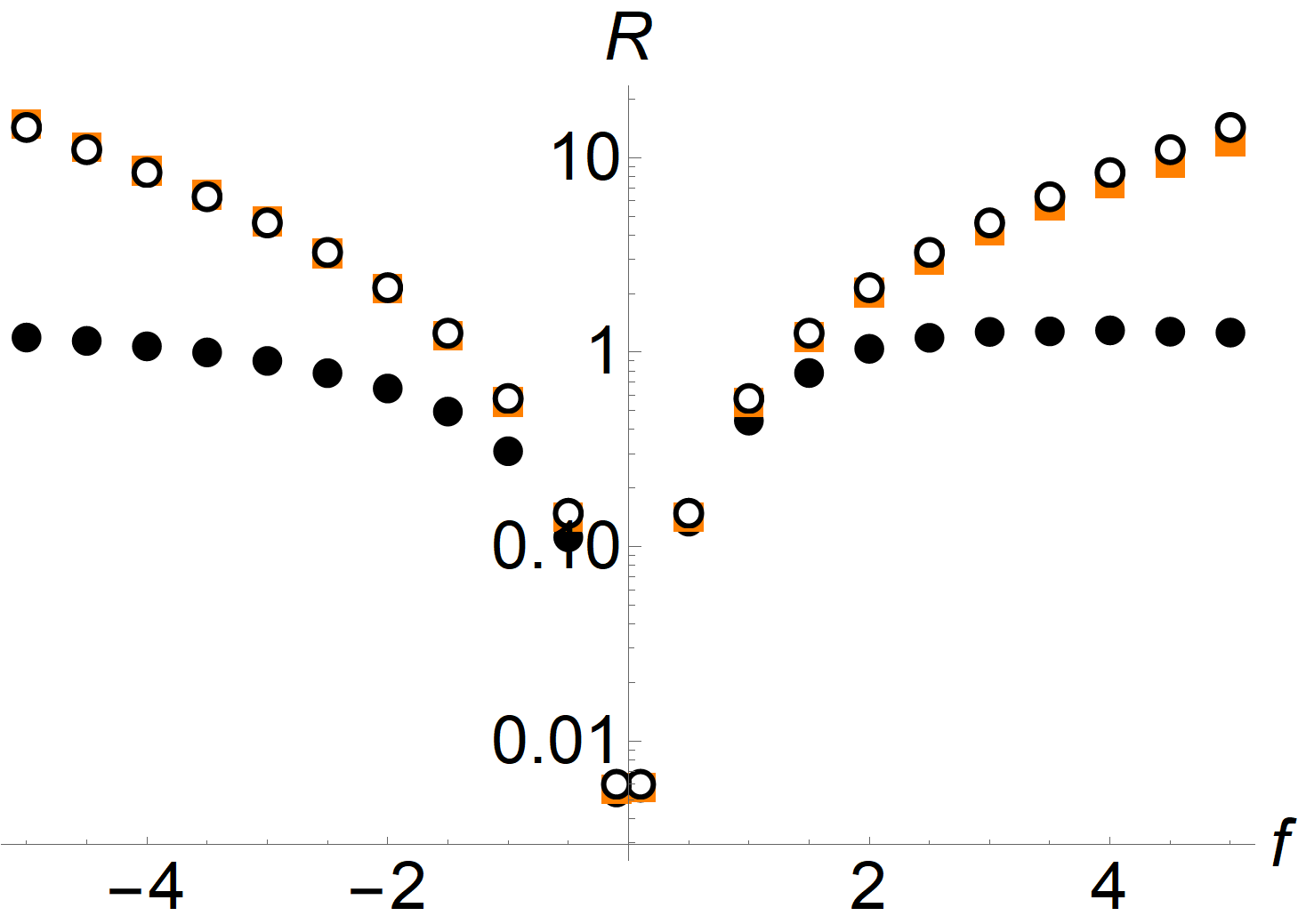}
\caption{The estimate on the pseudo entropy production from the TUR (filled black disks) and the CTUR (filled orange squares), compared to the pseudo entropy production $R_\text{L}$ of only the left ring (empty circles). \label{fig-two-cycle-estimate}}
\end{figure}
In the case of small bias, it is expected that the (pseudo) entropy production cannot be estimated accurately, since it mostly originates from the right ring, and is not reflected in the current through the left ring.
Thus, at most, we can estimate the contribution to the (pseudo) entropy production that originates from the left ring.
To illustrate this, we compare the estimate of $R$ obtained from the TUR and CTUR with the pseudo entropy production $R_\text{L}$ of only the left ring, i.~e.~the system in Fig.~\ref{fig-two-cycle} with $f_0 = 0$.
There is no rigorous inequality between the two, since the bias in the right ring may affect the occupation probabilities and currents in the left ring.
However, we do observe that the CTUR for the current in the left ring almost exactly reproduces $R_L$.
This is an example of a more general statement: While the CTUR can potentially provide an accurate estimate of the (pseudo) entropy production, it can only capture the contributions to entropy production that correspond to the chosen current observable.

\subsection{Transport in a two-dimensional channel} \label{sec-demonstration-channel}
\begin{figure}
\includegraphics[width=.49\textwidth]{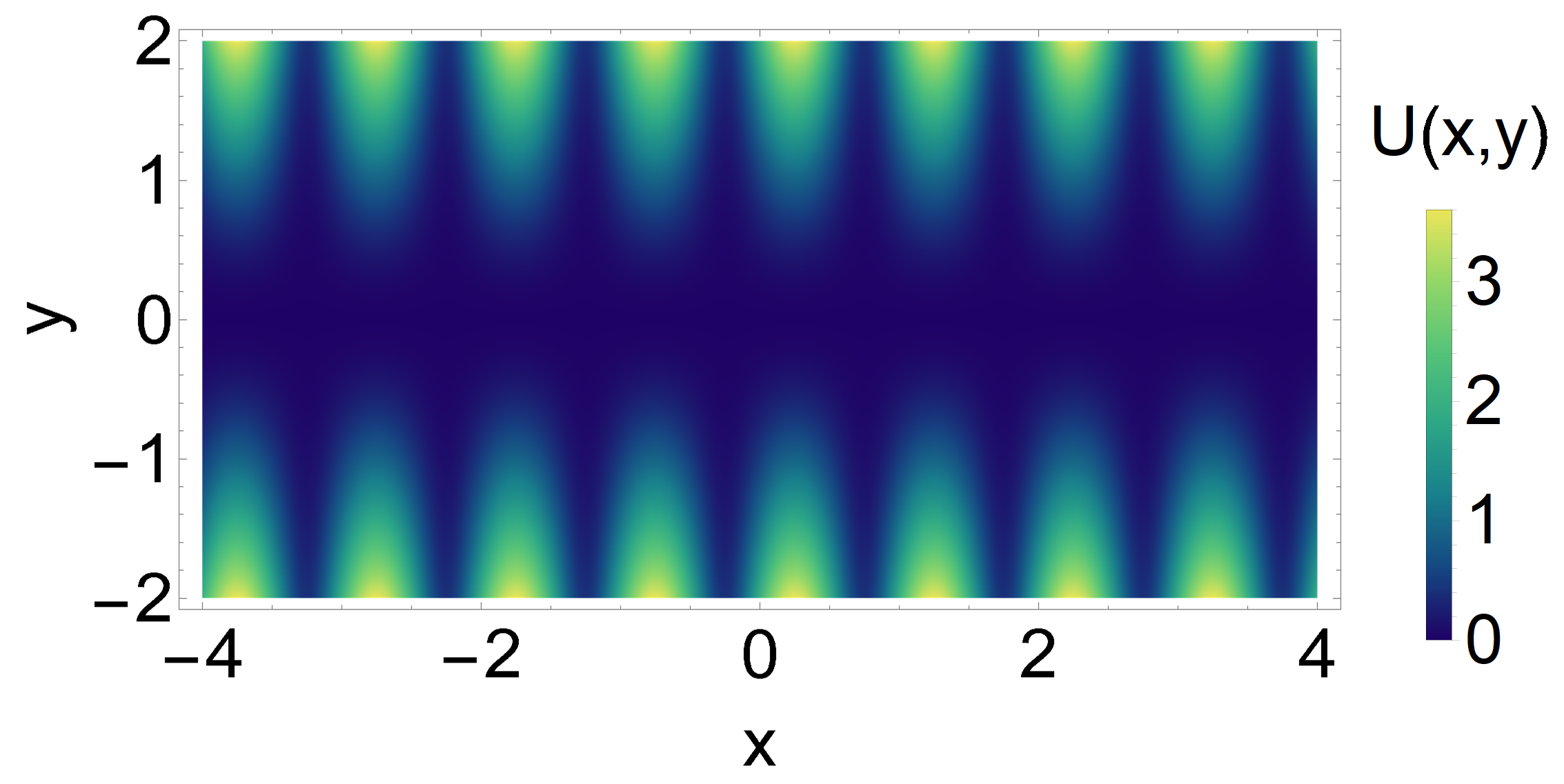}\\
\includegraphics[width=.49\textwidth]{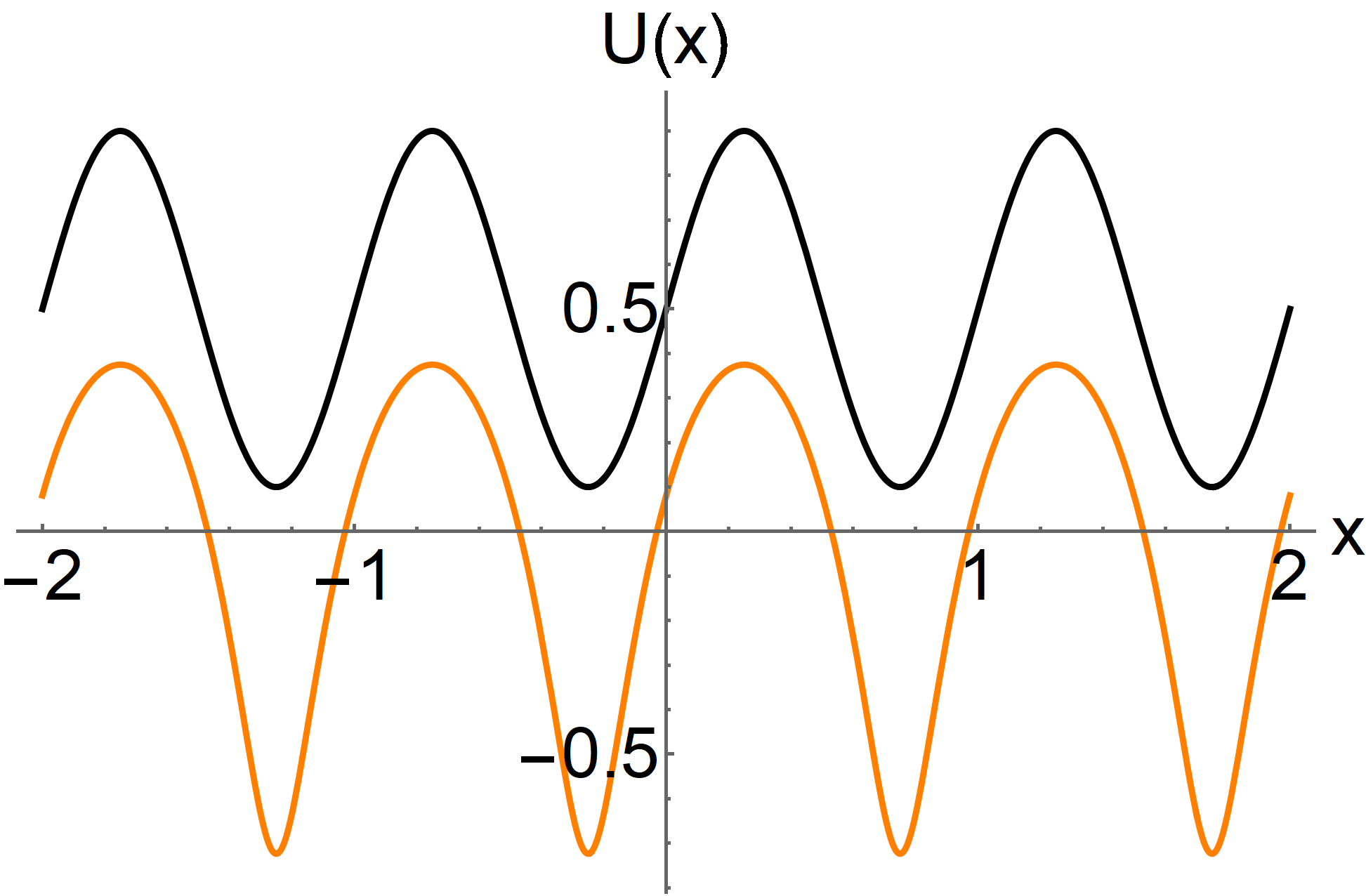}
\caption{The potential for the two-dimensional channel. The upper panel shows the two-dimensional potential \eqref{channel-potential}, the lower panel a section of the two-dimensional potential at $y = 1$ (black) and the effective one-dimensional potential \eqref{channel-potential-1d} (orange). The remaining parameters are $k = 1$, $L = 1$ and $\alpha = 0.8$.} \label{fig-channel-potential}
\end{figure}

\begin{figure}
\includegraphics[width=.49\textwidth]{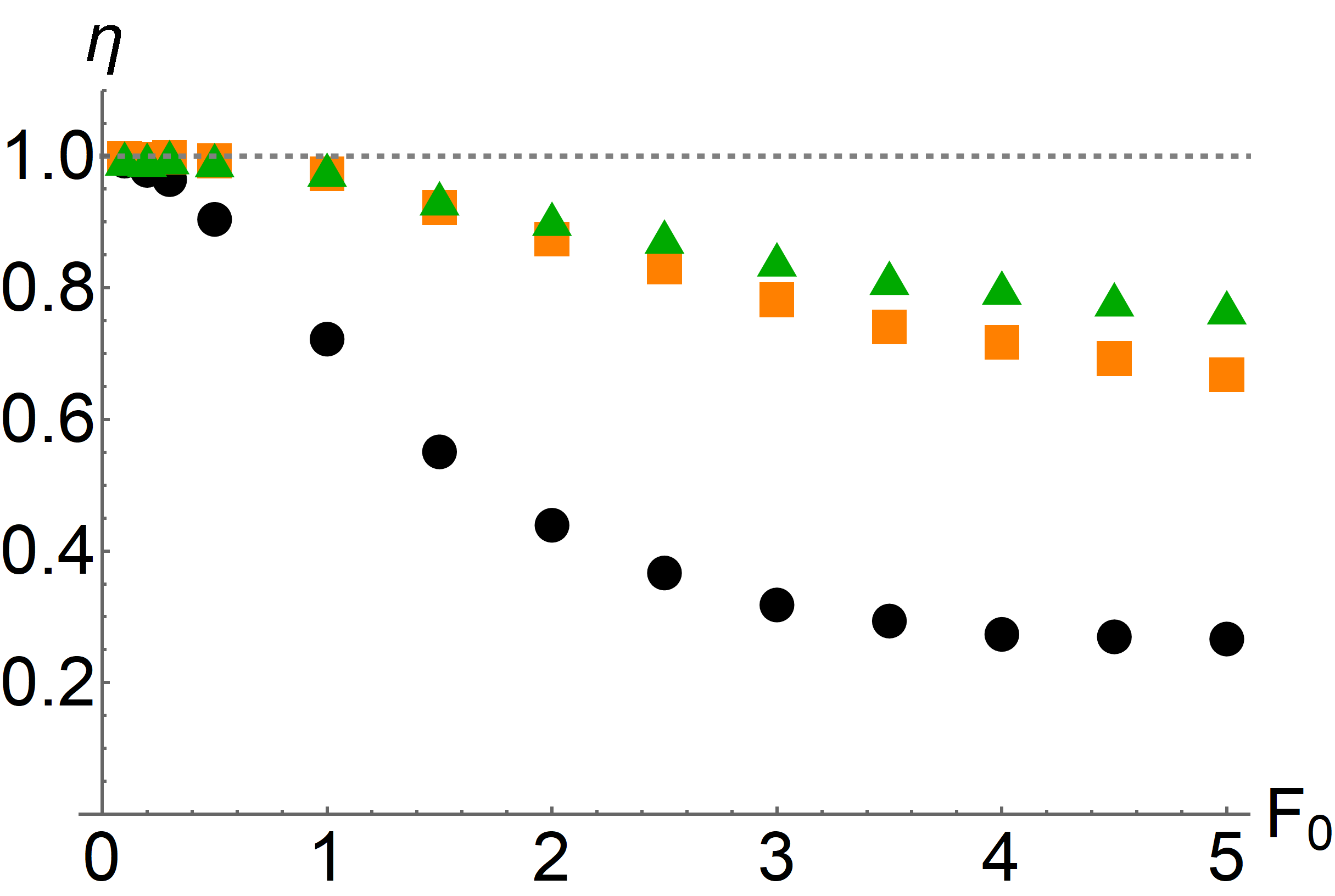}
\caption{The ratios defined in \eqref{transport-eff} for the current in the longitudinal direction in the channel potential \eqref{channel-potential} as a function of the bias force $F_0$. The black circles correspond to the TUR, the green triangles and orange squares to the CTUR with the approximate local mean velocity given by \eqref{channel-meanvel-1D} and \eqref{channel-meanvel-eq}, respectively. The data are obtained from Langevin simulations with $k = 1$, $L = 1$, $\alpha = 0.8$ and $\kb T = 0.5$ for $10^5$ trajectories. } \label{fig-channel-tur}
\end{figure}

Both models in the previous example have a very simple geometry: transitions only occur between neighboring sites along one-dimensional structures.
To illustrate the usefulness of the CTUR also in more complicated situations, we study the transport of a Brownian particle in a two-dimensional, soft-walled channel described by the potential
\begin{align}
U(x,y) = \frac{k}{2} \Big( 1 + \alpha \sin(\lambda x) \Big) y^2 \label{channel-potential} .
\end{align}
Along the transverse $y$-direction, the particle is confined in a parabolic trap, whose strength is periodically modulated in the longitudinal $x$-direction, see Fig.~\ref{fig-channel-potential}.
The period $L$ of the modulation is described by the wave number $\lambda = 2\pi/L$, its amplitude by the parameter $0 \leq \alpha < 1$, where the condition $\alpha < 1$ ensures that the potential remains trapping in the transverse direction.
The longitudinal motion in the potential \eqref{channel-potential} is qualitatively similar to the motion in a one-dimensional periodic potential.
This analogy can be made explicit by assuming that the relaxation in the transverse direction is fast compared to the longitudinal motion, so that the transverse coordinate is distributed according to the equilibrium distribution,
\begin{align}
p^\text{eq}(y \vert x) = \frac{1}{Z(x)} \exp \bigg[ - \frac{U(x,y)}{\kb T} \bigg] \label{channel-separation},
\end{align}
where $p^\text{eq}(y \vert x)$ is the conditional probability density of finding the particle at transverse coordinate $y$ given the longitudinal position $x$ and $Z(x)$ is the normalizing partition function.
This assumption is justified for a relatively narrow channel and in the absence of bias.
If it holds, we can describe the motion in the longitudinal direction via an effective one-dimensional Langevin dynamics
\begin{align}
\dot{x}(t) = - \mu \partial_x \tilde{U}(x) + \sqrt{2 \mu \kb T} \xi(t), \label{channel-effective-langevin}
\end{align}
with the potential
\begin{align}
\tilde{U}(x) = - \kb T \ln \bigg( \int_{-\infty}^\infty dy \ e^{-\frac{U(x,y)}{\kb T}} \bigg) .
\end{align}
For the form \eqref{channel-potential} of the potential, the integral can be evaluated explicitly and we find
\begin{align}
\tilde{U}(x) = \frac{\kb T}{2} \ln \Big( 1 + \alpha \sin(\lambda x) \Big) \label{channel-potential-1d} .
\end{align}
The effective potential governing the motion in the longitudinal direction appears only due to diffusion in the transverse direction---in the center of the channel, the potential is flat---and is thus proportional to the temperature, rather than the depth of the channel potential.
If we assume \eqref{channel-effective-langevin} to describe the dynamics in the longitudinal direction, then we can use it to compute the local mean velocity in the presence of a constant bias $F_0$,
\begin{align}
\tilde{\nu}(x) = \mu \kb T \big(1 - e^{- \frac{F_0 L}{T}} \big) \frac{e^{\frac{\tilde{U}(x)-F_0 x}{\kb T}}}{\int_{x}^{x+L} dz \ e^{\frac{\tilde{U}(z)-F_0 z}{\kb T}}} \label{channel-meanvel-1D}.
\end{align}
We remark that this expression is only an approximation, since, strictly speaking, the assumption of a transverse equilibrium distribution \eqref{channel-separation} breaks down in the presence of a bias.
Thus, the true local mean velocity has to be computed using \eqref{channel-potential} and depends on both the longitudinal and transverse coordinate.
However, the solution of the two-dimensional problem is considerably harder and does not permit a closed-form expression. 
Thus, we use \eqref{channel-meanvel-1D} as an approximate expression and the corresponding observable
\begin{align}
Z = \int_0^\tau dt \ \tilde{\nu}(x(t))
\end{align}
to compute the CTUR for the current given by the displacement in the longitudinal direction.
Note that, in principle, we may obtain even simpler approximations for the local mean velocity.
Since we expect the dynamics to be described by an effectively one-dimensional motion, the local mean velocity and probability density are related by $\nu^\text{st}(x) \propto 1/p^\text{st}(x)$, see the discussion in Section \ref{sec-data-occ}.
Provided that the bias is not too strong, we may further expect the probability density to be qualitatively similar to the equilibrium density.
Thus, a crude approximation of the local mean velocity is given by
\begin{align}
\hat{\nu}(x) = \exp \Bigg[\frac{U\big(x,\sqrt{\frac{\kb T}{k}}\big)}{\kb T} \Bigg] \label{channel-meanvel-eq} ,
\end{align}
where we evaluated the potential at $y = \sqrt{\kb T/k}$, which is a measure of the average distance of the particle from the center of the channel.
The resulting bounds \eqref{transport-eff} obtained from the TUR and CTUR are shown in Fig.~\ref{fig-channel-tur}.
For small bias and thus close to equilibrium, both the TUR and CTUR are close to an equality \cite{Mac18}.
As we drive the system further away from equilibrium by increasing the bias, the TUR ratio decreases sharply.
In this regime, the CTUR with either approximate local mean velocity yields a tighter bound than the TUR and remains close to an equality.
For even stronger bias, the equilibrium argument behind the approximate expressions \eqref{channel-meanvel-1D} and \eqref{channel-meanvel-eq} as well as the one-dimensional approximation become increasingly invalid, and thus the tightness of the CTUR likewise starts to decrease.
However, we stress that for moderately strong bias, the estimate on the entropy production from the CTUR with \eqref{channel-meanvel-1D} is around $80 \%$ of the true value, which is considerably tighter than the TUR estimate of less than $20 \%$.

\section{Discussion} \label{sec-discussion}
In this article, we have shown how to improve the TUR by taking into account the correlations between current and non-current observables and that the resulting inequality can yield a much improved estimate of entropy production.
We remark that \eqref{fri-corr} is more general: It allows us to improve any bound provided by the FRI, provided that we can find a quantity whose average is invariant under a suitable perturbation of the dynamics.
This is a manifestation of the monotonicity of information established in \eqref{information-monotonic}.
Since many generalizations of the TUR may be derived from the FRI \cite{Ter18,Koy19,Koy20,Van20,Liu20}, these generalizations can be improved in a completely analogous manner, by exploiting the existence of symmetries and conserved quantities.
For example, in Ref.~\cite{Koy19}, the TUR was generalized from steady states to systems in the presence of time-periodic driving.
In this case, the perturbation that leads to the TUR is a change in the driving frequency, and thus any non-current observable that is independent of the driving frequency can be used to obtain a tighter bound, similar to \eqref{TUR-corr}.

In the presence of several currents, it is only possible to estimate the partial entropy production corresponding to the measured current \cite{Pol16,Ots20}, see also the discussion in Section \ref{sec-demonstration-markov}. 
In such situations, measuring several currents and state-dependent observables is required to obtain a good estimate on the entropy production using \eqref{TUR-multi}.
We note that, since the derivation of Ref.~\cite{Pol16} involves minimizing the entropy production constrained on the measured current while keeping the steady state probability fixed, the CTUR extends in a straightforward manner to the tighter bounds involving the partial entropy production.
In general, caution is needed when interpreting the results of the estimation of entropy production for complex systems.
This is particularly true for biological systems, in which there are typically many out of equilibrium processes that to not directly contribute to any directly accessible observable, yet still result in (considerable) contributions to entropy production.
For example, observing the motion a biological cell can yield an accurate estimate on the entropy production corresponding to its mechanical motion, but it should not be interpreted as the entropy production of the cell, which involves a multitude of different chemical processes.

Finally, the fact that the model of $\text{F}_1$-ATPase is close to saturating the CTUR at biological conditions poses the question of whether this finding extends to models of other molecular motors.
In the light of interpreting \eqref{TUR-corr} as a transport efficiency, \eqref{transport-eff}, this appears reasonable:
Achieving precise transport at minimal dissipation would obviously be advantageous for any machine, whether artificial or naturally occurring.
We leave this issue for future research.

\begin{acknowledgments}
\textbf{Acknowledgments.} This work was supported by KAKENHI (Nos. 17H01148, 19H05795 and 20K20425).
\end{acknowledgments}


\begin{thebibliography}{36}%
\makeatletter
\providecommand \@ifxundefined [1]{%
 \@ifx{#1\undefined}
}%
\providecommand \@ifnum [1]{%
 \ifnum #1\expandafter \@firstoftwo
 \else \expandafter \@secondoftwo
 \fi
}%
\providecommand \@ifx [1]{%
 \ifx #1\expandafter \@firstoftwo
 \else \expandafter \@secondoftwo
 \fi
}%
\providecommand \natexlab [1]{#1}%
\providecommand \enquote  [1]{``#1''}%
\providecommand \bibnamefont  [1]{#1}%
\providecommand \bibfnamefont [1]{#1}%
\providecommand \citenamefont [1]{#1}%
\providecommand \href@noop [0]{\@secondoftwo}%
\providecommand \href [0]{\begingroup \@sanitize@url \@href}%
\providecommand \@href[1]{\@@startlink{#1}\@@href}%
\providecommand \@@href[1]{\endgroup#1\@@endlink}%
\providecommand \@sanitize@url [0]{\catcode `\\12\catcode `\$12\catcode
  `\&12\catcode `\#12\catcode `\^12\catcode `\_12\catcode `\%12\relax}%
\providecommand \@@startlink[1]{}%
\providecommand \@@endlink[0]{}%
\providecommand \url  [0]{\begingroup\@sanitize@url \@url }%
\providecommand \@url [1]{\endgroup\@href {#1}{\urlprefix }}%
\providecommand \urlprefix  [0]{URL }%
\providecommand \Eprint [0]{\href }%
\providecommand \doibase [0]{https://doi.org/}%
\providecommand \selectlanguage [0]{\@gobble}%
\providecommand \bibinfo  [0]{\@secondoftwo}%
\providecommand \bibfield  [0]{\@secondoftwo}%
\providecommand \translation [1]{[#1]}%
\providecommand \BibitemOpen [0]{}%
\providecommand \bibitemStop [0]{}%
\providecommand \bibitemNoStop [0]{.\EOS\space}%
\providecommand \EOS [0]{\spacefactor3000\relax}%
\providecommand \BibitemShut  [1]{\csname bibitem#1\endcsname}%
\let\auto@bib@innerbib\@empty
\bibitem [{\citenamefont {Sekimoto}(2010)}]{Sek10}%
  \BibitemOpen
  \bibfield  {author} {\bibinfo {author} {\bibfnamefont {K.}~\bibnamefont
  {Sekimoto}},\ }\href {https://books.google.de/books?id=8Fq7BQAAQBAJ} {\emph
  {\bibinfo {title} {Stochastic Energetics}}},\ Lecture Notes in Physics\
  (\bibinfo  {publisher} {Springer Berlin Heidelberg},\ \bibinfo {year}
  {2010})\BibitemShut {NoStop}%
\bibitem [{\citenamefont {Seifert}(2012)}]{Sei12}%
  \BibitemOpen
  \bibfield  {author} {\bibinfo {author} {\bibfnamefont {U.}~\bibnamefont
  {Seifert}},\ }\bibfield  {title} {\bibinfo {title} {Stochastic
  thermodynamics, fluctuation theorems and molecular machines},\ }\href
  {http://stacks.iop.org/0034-4885/75/i=12/a=126001} {\bibfield  {journal}
  {\bibinfo  {journal} {Rep. Prog. Phys.}\ }\textbf {\bibinfo {volume} {75}},\
  \bibinfo {pages} {126001} (\bibinfo {year} {2012})}\BibitemShut {NoStop}%
\bibitem [{\citenamefont {Harada}\ and\ \citenamefont {Sasa}(2005)}]{Har05}%
  \BibitemOpen
  \bibfield  {author} {\bibinfo {author} {\bibfnamefont {T.}~\bibnamefont
  {Harada}}\ and\ \bibinfo {author} {\bibfnamefont {S.-i.}\ \bibnamefont
  {Sasa}},\ }\bibfield  {title} {\bibinfo {title} {Equality connecting energy
  dissipation with a violation of the fluctuation-response relation},\ }\href
  {https://doi.org/10.1103/PhysRevLett.95.130602} {\bibfield  {journal}
  {\bibinfo  {journal} {Phys. Rev. Lett.}\ }\textbf {\bibinfo {volume} {95}},\
  \bibinfo {pages} {130602} (\bibinfo {year} {2005})}\BibitemShut {NoStop}%
\bibitem [{\citenamefont {Li}\ \emph {et~al.}(2019)\citenamefont {Li},
  \citenamefont {Horowitz}, \citenamefont {Gingrich},\ and\ \citenamefont
  {Fakhri}}]{Li19}%
  \BibitemOpen
  \bibfield  {author} {\bibinfo {author} {\bibfnamefont {J.}~\bibnamefont
  {Li}}, \bibinfo {author} {\bibfnamefont {J.~M.}\ \bibnamefont {Horowitz}},
  \bibinfo {author} {\bibfnamefont {T.~R.}\ \bibnamefont {Gingrich}},\ and\
  \bibinfo {author} {\bibfnamefont {N.}~\bibnamefont {Fakhri}},\ }\bibfield
  {title} {\bibinfo {title} {Quantifying dissipation using fluctuating
  currents},\ }\href@noop {} {\bibfield  {journal} {\bibinfo  {journal} {Nature
  Comm.}\ }\textbf {\bibinfo {volume} {10}},\ \bibinfo {pages} {1} (\bibinfo
  {year} {2019})}\BibitemShut {NoStop}%
\bibitem [{\citenamefont {Manikandan}\ \emph {et~al.}(2020)\citenamefont
  {Manikandan}, \citenamefont {Gupta},\ and\ \citenamefont
  {Krishnamurthy}}]{Man20}%
  \BibitemOpen
  \bibfield  {author} {\bibinfo {author} {\bibfnamefont {S.~K.}\ \bibnamefont
  {Manikandan}}, \bibinfo {author} {\bibfnamefont {D.}~\bibnamefont {Gupta}},\
  and\ \bibinfo {author} {\bibfnamefont {S.}~\bibnamefont {Krishnamurthy}},\
  }\bibfield  {title} {\bibinfo {title} {Inferring entropy production from
  short experiments},\ }\href {https://doi.org/10.1103/PhysRevLett.124.120603}
  {\bibfield  {journal} {\bibinfo  {journal} {Phys. Rev. Lett.}\ }\textbf
  {\bibinfo {volume} {124}},\ \bibinfo {pages} {120603} (\bibinfo {year}
  {2020})}\BibitemShut {NoStop}%
\bibitem [{\citenamefont {Otsubo}\ \emph {et~al.}(2020)\citenamefont {Otsubo},
  \citenamefont {Ito}, \citenamefont {Dechant},\ and\ \citenamefont
  {Sagawa}}]{Ots20}%
  \BibitemOpen
  \bibfield  {author} {\bibinfo {author} {\bibfnamefont {S.}~\bibnamefont
  {Otsubo}}, \bibinfo {author} {\bibfnamefont {S.}~\bibnamefont {Ito}},
  \bibinfo {author} {\bibfnamefont {A.}~\bibnamefont {Dechant}},\ and\ \bibinfo
  {author} {\bibfnamefont {T.}~\bibnamefont {Sagawa}},\ }\bibfield  {title}
  {\bibinfo {title} {Estimating entropy production by machine learning of
  short-time fluctuating currents},\ }\href
  {https://doi.org/10.1103/PhysRevE.101.062106} {\bibfield  {journal} {\bibinfo
   {journal} {Phys. Rev. E}\ }\textbf {\bibinfo {volume} {101}},\ \bibinfo
  {pages} {062106} (\bibinfo {year} {2020})}\BibitemShut {NoStop}%
\bibitem [{\citenamefont {Van~Vu}\ \emph {et~al.}(2020)\citenamefont {Van~Vu},
  \citenamefont {Vo},\ and\ \citenamefont {Hasegawa}}]{Vu20}%
  \BibitemOpen
  \bibfield  {author} {\bibinfo {author} {\bibfnamefont {T.}~\bibnamefont
  {Van~Vu}}, \bibinfo {author} {\bibfnamefont {V.~T.}\ \bibnamefont {Vo}},\
  and\ \bibinfo {author} {\bibfnamefont {Y.}~\bibnamefont {Hasegawa}},\
  }\bibfield  {title} {\bibinfo {title} {Entropy production estimation with
  optimal current},\ }\href {https://doi.org/10.1103/PhysRevE.101.042138}
  {\bibfield  {journal} {\bibinfo  {journal} {Phys. Rev. E}\ }\textbf {\bibinfo
  {volume} {101}},\ \bibinfo {pages} {042138} (\bibinfo {year}
  {2020})}\BibitemShut {NoStop}%
\bibitem [{\citenamefont {Horowitz}\ and\ \citenamefont
  {Gingrich}(2020)}]{Hor20}%
  \BibitemOpen
  \bibfield  {author} {\bibinfo {author} {\bibfnamefont {J.~M.}\ \bibnamefont
  {Horowitz}}\ and\ \bibinfo {author} {\bibfnamefont {T.~R.}\ \bibnamefont
  {Gingrich}},\ }\bibfield  {title} {\bibinfo {title} {Thermodynamic
  uncertainty relations constrain non-equilibrium fluctuations},\ }\href
  {https://doi.org/10.1038/s41567-019-0702-6} {\bibfield  {journal} {\bibinfo
  {journal} {Nature Phys.}\ }\textbf {\bibinfo {volume} {16}},\ \bibinfo
  {pages} {15} (\bibinfo {year} {2020})}\BibitemShut {NoStop}%
\bibitem [{\citenamefont {Barato}\ and\ \citenamefont {Seifert}(2015)}]{Bar15}%
  \BibitemOpen
  \bibfield  {author} {\bibinfo {author} {\bibfnamefont {A.~C.}\ \bibnamefont
  {Barato}}\ and\ \bibinfo {author} {\bibfnamefont {U.}~\bibnamefont
  {Seifert}},\ }\bibfield  {title} {\bibinfo {title} {Thermodynamic uncertainty
  relation for biomolecular processes},\ }\href
  {https://journals.aps.org/prl/abstract/10.1103/PhysRevLett.114.158101}
  {\bibfield  {journal} {\bibinfo  {journal} {Phys. Rev. Lett.}\ }\textbf
  {\bibinfo {volume} {114}},\ \bibinfo {pages} {158101} (\bibinfo {year}
  {2015})}\BibitemShut {NoStop}%
\bibitem [{\citenamefont {Gingrich}\ \emph {et~al.}(2016)\citenamefont
  {Gingrich}, \citenamefont {Horowitz}, \citenamefont {Perunov},\ and\
  \citenamefont {England}}]{Gin16}%
  \BibitemOpen
  \bibfield  {author} {\bibinfo {author} {\bibfnamefont {T.~R.}\ \bibnamefont
  {Gingrich}}, \bibinfo {author} {\bibfnamefont {J.~M.}\ \bibnamefont
  {Horowitz}}, \bibinfo {author} {\bibfnamefont {N.}~\bibnamefont {Perunov}},\
  and\ \bibinfo {author} {\bibfnamefont {J.~L.}\ \bibnamefont {England}},\
  }\bibfield  {title} {\bibinfo {title} {Dissipation bounds all steady-state
  current fluctuations},\ }\href
  {https://doi.org/10.1103/PhysRevLett.116.120601} {\bibfield  {journal}
  {\bibinfo  {journal} {Phys. Rev. Lett.}\ }\textbf {\bibinfo {volume} {116}},\
  \bibinfo {pages} {120601} (\bibinfo {year} {2016})}\BibitemShut {NoStop}%
\bibitem [{\citenamefont {Dechant}\ and\ \citenamefont
  {Sasa}(2018{\natexlab{a}})}]{Dec17}%
  \BibitemOpen
  \bibfield  {author} {\bibinfo {author} {\bibfnamefont {A.}~\bibnamefont
  {Dechant}}\ and\ \bibinfo {author} {\bibfnamefont {S.-i.}\ \bibnamefont
  {Sasa}},\ }\bibfield  {title} {\bibinfo {title} {{Current fluctuations and
  transport efficiency for general Langevin systems}},\ }\href
  {http://stacks.iop.org/1742-5468/2018/i=6/a=063209} {\bibfield  {journal}
  {\bibinfo  {journal} {J. Stat. Mech. Theory E.}\ }\textbf {\bibinfo {volume}
  {2018}},\ \bibinfo {pages} {063209} (\bibinfo {year}
  {2018}{\natexlab{a}})}\BibitemShut {NoStop}%
\bibitem [{\citenamefont {Pietzonka}\ \emph {et~al.}(2017)\citenamefont
  {Pietzonka}, \citenamefont {Ritort},\ and\ \citenamefont {Seifert}}]{Pie17}%
  \BibitemOpen
  \bibfield  {author} {\bibinfo {author} {\bibfnamefont {P.}~\bibnamefont
  {Pietzonka}}, \bibinfo {author} {\bibfnamefont {F.}~\bibnamefont {Ritort}},\
  and\ \bibinfo {author} {\bibfnamefont {U.}~\bibnamefont {Seifert}},\
  }\bibfield  {title} {\bibinfo {title} {Finite-time generalization of the
  thermodynamic uncertainty relation},\ }\href
  {https://doi.org/10.1103/PhysRevE.96.012101} {\bibfield  {journal} {\bibinfo
  {journal} {Phys. Rev. E}\ }\textbf {\bibinfo {volume} {96}},\ \bibinfo
  {pages} {012101} (\bibinfo {year} {2017})}\BibitemShut {NoStop}%
\bibitem [{\citenamefont {Dechant}\ and\ \citenamefont
  {i.~Sasa}(2020)}]{Dec20b}%
  \BibitemOpen
  \bibfield  {author} {\bibinfo {author} {\bibfnamefont {A.}~\bibnamefont
  {Dechant}}\ and\ \bibinfo {author} {\bibfnamefont {S.}~\bibnamefont
  {i.~Sasa}},\ }\href@noop {} {\bibinfo {title} {Continuous time-reversal and
  equality in the thermodynamic uncertainty relation}} (\bibinfo {year}
  {2020}),\ \Eprint {https://arxiv.org/abs/2010.14769} {arXiv:2010.14769
  [cond-mat.stat-mech]} \BibitemShut {NoStop}%
\bibitem [{\citenamefont {Hwang}\ and\ \citenamefont {Hyeon}(2018)}]{Hwa18}%
  \BibitemOpen
  \bibfield  {author} {\bibinfo {author} {\bibfnamefont {W.}~\bibnamefont
  {Hwang}}\ and\ \bibinfo {author} {\bibfnamefont {C.}~\bibnamefont {Hyeon}},\
  }\bibfield  {title} {\bibinfo {title} {Energetic costs, precision, and
  transport efficiency of molecular motors},\ }\href@noop {} {\bibfield
  {journal} {\bibinfo  {journal} {J. Phys. Chem. Lett.}\ }\textbf {\bibinfo
  {volume} {9}},\ \bibinfo {pages} {513} (\bibinfo {year} {2018})}\BibitemShut
  {NoStop}%
\bibitem [{\citenamefont {Dechant}\ and\ \citenamefont
  {Sasa}(2018{\natexlab{b}})}]{Dec18}%
  \BibitemOpen
  \bibfield  {author} {\bibinfo {author} {\bibfnamefont {A.}~\bibnamefont
  {Dechant}}\ and\ \bibinfo {author} {\bibfnamefont {S.-i.}\ \bibnamefont
  {Sasa}},\ }\bibfield  {title} {\bibinfo {title} {Entropic bounds on currents
  in {Langevin} systems},\ }\href {https://doi.org/10.1103/PhysRevE.97.062101}
  {\bibfield  {journal} {\bibinfo  {journal} {Phys. Rev. E}\ }\textbf {\bibinfo
  {volume} {97}},\ \bibinfo {pages} {062101} (\bibinfo {year}
  {2018}{\natexlab{b}})}\BibitemShut {NoStop}%
\bibitem [{\citenamefont {Pal}\ \emph {et~al.}(2020)\citenamefont {Pal},
  \citenamefont {Saryal}, \citenamefont {Segal}, \citenamefont {Mahesh},\ and\
  \citenamefont {Agarwalla}}]{Pal20}%
  \BibitemOpen
  \bibfield  {author} {\bibinfo {author} {\bibfnamefont {S.}~\bibnamefont
  {Pal}}, \bibinfo {author} {\bibfnamefont {S.}~\bibnamefont {Saryal}},
  \bibinfo {author} {\bibfnamefont {D.}~\bibnamefont {Segal}}, \bibinfo
  {author} {\bibfnamefont {T.~S.}\ \bibnamefont {Mahesh}},\ and\ \bibinfo
  {author} {\bibfnamefont {B.~K.}\ \bibnamefont {Agarwalla}},\ }\bibfield
  {title} {\bibinfo {title} {Experimental study of the thermodynamic
  uncertainty relation},\ }\href
  {https://doi.org/10.1103/PhysRevResearch.2.022044} {\bibfield  {journal}
  {\bibinfo  {journal} {Phys. Rev. Research}\ }\textbf {\bibinfo {volume}
  {2}},\ \bibinfo {pages} {022044(R)} (\bibinfo {year} {2020})}\BibitemShut
  {NoStop}%
\bibitem [{\citenamefont {Liu}\ \emph {et~al.}(2020)\citenamefont {Liu},
  \citenamefont {Gong},\ and\ \citenamefont {Ueda}}]{Liu20}%
  \BibitemOpen
  \bibfield  {author} {\bibinfo {author} {\bibfnamefont {K.}~\bibnamefont
  {Liu}}, \bibinfo {author} {\bibfnamefont {Z.}~\bibnamefont {Gong}},\ and\
  \bibinfo {author} {\bibfnamefont {M.}~\bibnamefont {Ueda}},\ }\bibfield
  {title} {\bibinfo {title} {Thermodynamic uncertainty relation for arbitrary
  initial states},\ }\href {https://doi.org/10.1103/PhysRevLett.125.140602}
  {\bibfield  {journal} {\bibinfo  {journal} {Phys. Rev. Lett.}\ }\textbf
  {\bibinfo {volume} {125}},\ \bibinfo {pages} {140602} (\bibinfo {year}
  {2020})}\BibitemShut {NoStop}%
\bibitem [{\citenamefont {Koyuk}\ and\ \citenamefont {Seifert}(2019)}]{Koy19}%
  \BibitemOpen
  \bibfield  {author} {\bibinfo {author} {\bibfnamefont {T.}~\bibnamefont
  {Koyuk}}\ and\ \bibinfo {author} {\bibfnamefont {U.}~\bibnamefont
  {Seifert}},\ }\bibfield  {title} {\bibinfo {title} {Operationally accessible
  bounds on fluctuations and entropy production in periodically driven
  systems},\ }\href {https://doi.org/10.1103/PhysRevLett.122.230601} {\bibfield
   {journal} {\bibinfo  {journal} {Phys. Rev. Lett.}\ }\textbf {\bibinfo
  {volume} {122}},\ \bibinfo {pages} {230601} (\bibinfo {year}
  {2019})}\BibitemShut {NoStop}%
\bibitem [{\citenamefont {Koyuk}\ and\ \citenamefont {Seifert}(2020)}]{Koy20}%
  \BibitemOpen
  \bibfield  {author} {\bibinfo {author} {\bibfnamefont {T.}~\bibnamefont
  {Koyuk}}\ and\ \bibinfo {author} {\bibfnamefont {U.}~\bibnamefont
  {Seifert}},\ }\bibfield  {title} {\bibinfo {title} {Thermodynamic uncertainty
  relation for time-dependent driving},\ }\href
  {https://doi.org/10.1103/PhysRevLett.125.260604} {\bibfield  {journal}
  {\bibinfo  {journal} {Phys. Rev. Lett.}\ }\textbf {\bibinfo {volume} {125}},\
  \bibinfo {pages} {260604} (\bibinfo {year} {2020})}\BibitemShut {NoStop}%
\bibitem [{\citenamefont {Van~Vu}\ and\ \citenamefont
  {Hasegawa}(2020)}]{Van20}%
  \BibitemOpen
  \bibfield  {author} {\bibinfo {author} {\bibfnamefont {T.}~\bibnamefont
  {Van~Vu}}\ and\ \bibinfo {author} {\bibfnamefont {Y.}~\bibnamefont
  {Hasegawa}},\ }\bibfield  {title} {\bibinfo {title} {Thermodynamic
  uncertainty relations under arbitrary control protocols},\ }\href
  {https://doi.org/10.1103/PhysRevResearch.2.013060} {\bibfield  {journal}
  {\bibinfo  {journal} {Phys. Rev. Research}\ }\textbf {\bibinfo {volume}
  {2}},\ \bibinfo {pages} {013060} (\bibinfo {year} {2020})}\BibitemShut
  {NoStop}%
\bibitem [{\citenamefont {Zimmermann}\ and\ \citenamefont
  {Seifert}(2012)}]{Zim12}%
  \BibitemOpen
  \bibfield  {author} {\bibinfo {author} {\bibfnamefont {E.}~\bibnamefont
  {Zimmermann}}\ and\ \bibinfo {author} {\bibfnamefont {U.}~\bibnamefont
  {Seifert}},\ }\bibfield  {title} {\bibinfo {title} {Efficiencies of a
  molecular motor: a generic hybrid model applied to the {F1-ATPase}},\
  }\href@noop {} {\bibfield  {journal} {\bibinfo  {journal} {New J. Phys.}\
  }\textbf {\bibinfo {volume} {14}},\ \bibinfo {pages} {103023} (\bibinfo
  {year} {2012})}\BibitemShut {NoStop}%
\bibitem [{\citenamefont {Kawaguchi}\ \emph {et~al.}(2014)\citenamefont
  {Kawaguchi}, \citenamefont {i.~Sasa},\ and\ \citenamefont {Sagawa}}]{Kaw14}%
  \BibitemOpen
  \bibfield  {author} {\bibinfo {author} {\bibfnamefont {K.}~\bibnamefont
  {Kawaguchi}}, \bibinfo {author} {\bibfnamefont {S.}~\bibnamefont {i.~Sasa}},\
  and\ \bibinfo {author} {\bibfnamefont {T.}~\bibnamefont {Sagawa}},\
  }\bibfield  {title} {\bibinfo {title} {Nonequilibrium dissipation-free
  transport in f1-atpase and the thermodynamic role of asymmetric
  allosterism},\ }\href
  {https://doi.org/https://doi.org/10.1016/j.bpj.2014.04.034} {\bibfield
  {journal} {\bibinfo  {journal} {Biophys. J.}\ }\textbf {\bibinfo {volume}
  {106}},\ \bibinfo {pages} {2450 } (\bibinfo {year} {2014})}\BibitemShut
  {NoStop}%
\bibitem [{\citenamefont {Dechant}\ and\ \citenamefont {Sasa}(2020)}]{Dec20}%
  \BibitemOpen
  \bibfield  {author} {\bibinfo {author} {\bibfnamefont {A.}~\bibnamefont
  {Dechant}}\ and\ \bibinfo {author} {\bibfnamefont {S.-i.}\ \bibnamefont
  {Sasa}},\ }\bibfield  {title} {\bibinfo {title}
  {Fluctuation{\textendash}response inequality out of equilibrium},\ }\href
  {https://doi.org/10.1073/pnas.1918386117} {\bibfield  {journal} {\bibinfo
  {journal} {Proc. Natl. Acad. Sci.}\ }\textbf {\bibinfo {volume} {117}},\
  \bibinfo {pages} {6430} (\bibinfo {year} {2020})}\BibitemShut {NoStop}%
\bibitem [{\citenamefont {Dechant}(2018)}]{Dec18c}%
  \BibitemOpen
  \bibfield  {author} {\bibinfo {author} {\bibfnamefont {A.}~\bibnamefont
  {Dechant}},\ }\bibfield  {title} {\bibinfo {title} {Multidimensional
  thermodynamic uncertainty relations},\ }\href
  {https://doi.org/10.1088/1751-8121/aaf3ff} {\bibfield  {journal} {\bibinfo
  {journal} {J. Phys. A Math. Theor.}\ }\textbf {\bibinfo {volume} {52}},\
  \bibinfo {pages} {035001} (\bibinfo {year} {2018})}\BibitemShut {NoStop}%
\bibitem [{\citenamefont {Radhakrishna~Rao}(1945)}]{Rao45}%
  \BibitemOpen
  \bibfield  {author} {\bibinfo {author} {\bibfnamefont {C.}~\bibnamefont
  {Radhakrishna~Rao}},\ }\bibfield  {title} {\bibinfo {title} {Information and
  the accuracy attainable in the estimation of statistical parameters},\
  }\href@noop {} {\bibfield  {journal} {\bibinfo  {journal} {Bull. Calcutta
  Math. Soc.}\ }\textbf {\bibinfo {volume} {37}},\ \bibinfo {pages} {81}
  (\bibinfo {year} {1945})}\BibitemShut {NoStop}%
\bibitem [{\citenamefont {Cram{\'e}r}(2016)}]{Cra16}%
  \BibitemOpen
  \bibfield  {author} {\bibinfo {author} {\bibfnamefont {H.}~\bibnamefont
  {Cram{\'e}r}},\ }\href@noop {} {\emph {\bibinfo {title} {Mathematical methods
  of statistics}}},\ Vol.~\bibinfo {volume} {9}\ (\bibinfo  {publisher}
  {Princeton university press},\ \bibinfo {year} {2016})\BibitemShut {NoStop}%
\bibitem [{\citenamefont {Chetrite}\ and\ \citenamefont
  {Touchette}(2015)}]{Che15}%
  \BibitemOpen
  \bibfield  {author} {\bibinfo {author} {\bibfnamefont {R.}~\bibnamefont
  {Chetrite}}\ and\ \bibinfo {author} {\bibfnamefont {H.}~\bibnamefont
  {Touchette}},\ }\bibfield  {title} {\bibinfo {title} {{Nonequilibrium Markov
  processes conditioned on large deviations}},\ }\href@noop {} {\bibfield
  {journal} {\bibinfo  {journal} {Ann. Henri Poincar{\'e}}\ }\textbf {\bibinfo
  {volume} {16}},\ \bibinfo {pages} {2005} (\bibinfo {year}
  {2015})}\BibitemShut {NoStop}%
\bibitem [{\citenamefont {Sasa}(2014)}]{Sas14}%
  \BibitemOpen
  \bibfield  {author} {\bibinfo {author} {\bibfnamefont {S.-i.}\ \bibnamefont
  {Sasa}},\ }\bibfield  {title} {\bibinfo {title} {Possible extended forms of
  thermodynamic entropy},\ }\href
  {https://doi.org/10.1088/1742-5468/2014/01/p01004} {\bibfield  {journal}
  {\bibinfo  {journal} {J. Stat. Mech. Theory E.}\ }\textbf {\bibinfo {volume}
  {2014}},\ \bibinfo {pages} {P01004} (\bibinfo {year} {2014})}\BibitemShut
  {NoStop}%
\bibitem [{\citenamefont {Hasegawa}\ and\ \citenamefont
  {Van~Vu}(2019)}]{Has19b}%
  \BibitemOpen
  \bibfield  {author} {\bibinfo {author} {\bibfnamefont {Y.}~\bibnamefont
  {Hasegawa}}\ and\ \bibinfo {author} {\bibfnamefont {T.}~\bibnamefont
  {Van~Vu}},\ }\bibfield  {title} {\bibinfo {title} {Uncertainty relations in
  stochastic processes: An information inequality approach},\ }\href@noop {}
  {\bibfield  {journal} {\bibinfo  {journal} {Phys. Rev. E}\ }\textbf {\bibinfo
  {volume} {99}},\ \bibinfo {pages} {062126} (\bibinfo {year}
  {2019})}\BibitemShut {NoStop}%
\bibitem [{\citenamefont {Pigolotti}\ \emph {et~al.}(2017)\citenamefont
  {Pigolotti}, \citenamefont {Neri}, \citenamefont {Rold\'an},\ and\
  \citenamefont {J\"ulicher}}]{Pig17}%
  \BibitemOpen
  \bibfield  {author} {\bibinfo {author} {\bibfnamefont {S.}~\bibnamefont
  {Pigolotti}}, \bibinfo {author} {\bibfnamefont {I.}~\bibnamefont {Neri}},
  \bibinfo {author} {\bibfnamefont {E.}~\bibnamefont {Rold\'an}},\ and\
  \bibinfo {author} {\bibfnamefont {F.}~\bibnamefont {J\"ulicher}},\ }\bibfield
   {title} {\bibinfo {title} {Generic properties of stochastic entropy
  production},\ }\href {https://doi.org/10.1103/PhysRevLett.119.140604}
  {\bibfield  {journal} {\bibinfo  {journal} {Phys. Rev. Lett.}\ }\textbf
  {\bibinfo {volume} {119}},\ \bibinfo {pages} {140604} (\bibinfo {year}
  {2017})}\BibitemShut {NoStop}%
\bibitem [{\citenamefont {Shiraishi}(2021)}]{Shi21}%
  \BibitemOpen
  \bibfield  {author} {\bibinfo {author} {\bibfnamefont {N.}~\bibnamefont
  {Shiraishi}},\ }\href@noop {} {\bibinfo {title} {Optimal thermodynamic
  uncertainty relation in markov jump processes}} (\bibinfo {year} {2021}),\
  \Eprint {https://arxiv.org/abs/2106.11634} {arXiv:2106.11634
  [cond-mat.stat-mech]} \BibitemShut {NoStop}%
\bibitem [{\citenamefont {Pal}\ \emph {et~al.}(2021)\citenamefont {Pal},
  \citenamefont {Reuveni},\ and\ \citenamefont {Rahav}}]{Pal21}%
  \BibitemOpen
  \bibfield  {author} {\bibinfo {author} {\bibfnamefont {A.}~\bibnamefont
  {Pal}}, \bibinfo {author} {\bibfnamefont {S.}~\bibnamefont {Reuveni}},\ and\
  \bibinfo {author} {\bibfnamefont {S.}~\bibnamefont {Rahav}},\ }\bibfield
  {title} {\bibinfo {title} {Thermodynamic uncertainty relation for systems
  with unidirectional transitions},\ }\href@noop {} {\bibfield  {journal}
  {\bibinfo  {journal} {Phys. Rev. Research}\ }\textbf {\bibinfo {volume}
  {3}},\ \bibinfo {pages} {013273} (\bibinfo {year} {2021})}\BibitemShut
  {NoStop}%
\bibitem [{\citenamefont {Macieszczak}\ \emph {et~al.}(2018)\citenamefont
  {Macieszczak}, \citenamefont {Brandner},\ and\ \citenamefont
  {Garrahan}}]{Mac18}%
  \BibitemOpen
  \bibfield  {author} {\bibinfo {author} {\bibfnamefont {K.}~\bibnamefont
  {Macieszczak}}, \bibinfo {author} {\bibfnamefont {K.}~\bibnamefont
  {Brandner}},\ and\ \bibinfo {author} {\bibfnamefont {J.~P.}\ \bibnamefont
  {Garrahan}},\ }\bibfield  {title} {\bibinfo {title} {Unified thermodynamic
  uncertainty relations in linear response},\ }\href
  {https://doi.org/10.1103/PhysRevLett.121.130601} {\bibfield  {journal}
  {\bibinfo  {journal} {Phys. Rev. Lett.}\ }\textbf {\bibinfo {volume} {121}},\
  \bibinfo {pages} {130601} (\bibinfo {year} {2018})}\BibitemShut {NoStop}%
\bibitem [{\citenamefont {Liepelt}\ and\ \citenamefont
  {Lipowsky}(2007)}]{Lie07}%
  \BibitemOpen
  \bibfield  {author} {\bibinfo {author} {\bibfnamefont {S.}~\bibnamefont
  {Liepelt}}\ and\ \bibinfo {author} {\bibfnamefont {R.}~\bibnamefont
  {Lipowsky}},\ }\bibfield  {title} {\bibinfo {title} {Kinesin's network of
  chemomechanical motor cycles},\ }\href
  {https://doi.org/10.1103/PhysRevLett.98.258102} {\bibfield  {journal}
  {\bibinfo  {journal} {Phys. Rev. Lett.}\ }\textbf {\bibinfo {volume} {98}},\
  \bibinfo {pages} {258102} (\bibinfo {year} {2007})}\BibitemShut {NoStop}%
\bibitem [{\citenamefont {Di~Terlizzi}\ and\ \citenamefont
  {Baiesi}(2018)}]{Ter18}%
  \BibitemOpen
  \bibfield  {author} {\bibinfo {author} {\bibfnamefont {I.}~\bibnamefont
  {Di~Terlizzi}}\ and\ \bibinfo {author} {\bibfnamefont {M.}~\bibnamefont
  {Baiesi}},\ }\bibfield  {title} {\bibinfo {title} {Kinetic uncertainty
  relation},\ }\href@noop {} {\bibfield  {journal} {\bibinfo  {journal} {J.
  Phys. A Math. Theor.}\ }\textbf {\bibinfo {volume} {52}},\ \bibinfo {pages}
  {02LT03} (\bibinfo {year} {2018})}\BibitemShut {NoStop}%
\bibitem [{\citenamefont {Polettini}\ \emph {et~al.}(2016)\citenamefont
  {Polettini}, \citenamefont {Lazarescu},\ and\ \citenamefont
  {Esposito}}]{Pol16}%
  \BibitemOpen
  \bibfield  {author} {\bibinfo {author} {\bibfnamefont {M.}~\bibnamefont
  {Polettini}}, \bibinfo {author} {\bibfnamefont {A.}~\bibnamefont
  {Lazarescu}},\ and\ \bibinfo {author} {\bibfnamefont {M.}~\bibnamefont
  {Esposito}},\ }\bibfield  {title} {\bibinfo {title} {Tightening the
  uncertainty principle for stochastic currents},\ }\href@noop {} {\bibfield
  {journal} {\bibinfo  {journal} {Phys. Rev. E}\ }\textbf {\bibinfo {volume}
  {94}},\ \bibinfo {pages} {052104} (\bibinfo {year} {2016})}\BibitemShut
  {NoStop}%
\end{thebibliography}

%

\appendix

\onecolumngrid

\section{Optimal state dependent observables for stochastic entropy production} \label{app-optimal-obs}
In Ref.~\cite{Dec20b}, an explicit expression for the variance of a stochastic current in a diffusion process was derived,
\begin{align}
\text{Var}_J &= \int_0^\tau dt \int_0^\tau ds \int d\bm{x} \int d\bm{y} \ \mu(\bm{x}) \mu(\bm{y}) \Big( p(\bm{x},t ; \bm{y},s) - p^\text{st}(\bm{x}) p^\text{st}(\bm{y}) \Big) \label{current-variance} \\
&\quad + \int_0^\tau dt \int_0^t ds \int d\bm{x} \int d\bm{y} \ \Big(\mu(\bm{y}) \psi(\bm{x}) - \mu(\bm{x}) \psi(\bm{y}) \Big) p(\bm{x},t ; \bm{y},s) \nn
&\quad + 2 \tau \int d\bm{x} \ \bm{w}(\bm{x}) \cdot \bm{B} \bm{w}(\bm{x}) p^\text{st}(\bm{x}) - \int_0^\tau dt \int_0^\tau ds \int d\bm{x} \int d\bm{y} \ \psi(\bm{x}) \psi(\bm{y}) p(\bm{x},t ; \bm{y},s) \n ,
\end{align}
where $p(\bm{x},t; \bm{y},s)$ is the joint probability density corresponding to \eqref{langevin} in the absence of the discrete degree of freedom.
In the above expression, we defined
\begin{align}
\mu(\bm{x}) = \bm{w}^\text{T}(\bm{x}) \nu^\text{st}(\bm{x}) \qquad \text{and} \qquad \psi(\bm{x}) = \bm{w}^\text{T}(\bm{x}) \bm{\phi}^\text{st}(\bm{x}) + \text{tr}\big(\bm{B} \bm{\mathcal{J}}_w(\bm{x}) \big).
\end{align}
Here, $\bm{\phi}^\text{st}(\bm{x})$ is the reversible part of the drift,
\begin{align}
\bm{\phi}^\text{st}(\bm{x}) = \bm{B} \grad \ln p^\text{st}(\bm{x}),
\end{align}
tr denotes the trace and $\bm{\mathcal{J}}_w(\bm{x})$ is the Jacobian matrix of the weighting function $\bm{w}(\bm{x})$.
In a similar manner, we may derive the covariance between the current $J$ and the state-dependent observable $Z$,
\begin{align}
\text{Cov}_{J,Z} &= \int_0^\tau dt \int_0^\tau ds \int d\bm{x} \int d\bm{y} \ z(\bm{x}) \mu(\bm{y}) \Big( p(\bm{x},t ; \bm{y},s) - p^\text{st}(\bm{x}) p^\text{st}(\bm{y}) \Big) \label{current-covariance} \\
&\quad + \int_0^\tau dt \int_0^t ds \int d\bm{x} \int d\bm{y} \ \Big(z(\bm{y}) \psi(\bm{x}) - z(\bm{x}) \psi(\bm{y}) \Big) p(\bm{x},t ; \bm{y},s) \n ,
\end{align}
while the variance of $Z$ is given by
\begin{align}
\text{Var}_Z = \int_0^\tau \int_0^\tau dt \int_0^\tau ds \int d\bm{x} \int d\bm{y} \ z(\bm{x}) z(\bm{y}) \Big( p(\bm{x},t ; \bm{y},s) - p^\text{st}(\bm{x}) p^\text{st}(\bm{y}) \Big) .
\end{align}
The stochastic entropy production $\Sigma$ is the current with the weighting function $\bm{w}(\bm{x}) = \bm{B}^{-1} \bm{\nu}^\text{st}(\bm{x})$.
For this choice, we find
\begin{align}
\psi(\bm{x}) = \bm{\nu}^\text{st,T}(\bm{x}) \grad \ln p^\text{st}(\bm{x}) + \grad^\text{T} \bm{\nu}^\text{st}(\bm{x}) = \frac{1}{p^\text{st}(\bm{x})} \Big( \grad^\text{T} \big( \bm{\nu}^\text{st}(\bm{x}) p^\text{st}(\bm{x}) \big) \Big) = 0 ,
\end{align}
where we used that the expression in parentheses is nothing but the steady-state condition of the Fokker-Planck equation \eqref{fpe}.
In this case, the expressions \eqref{current-variance} and \eqref{current-covariance} simplify,
\begin{align}
\text{Var}_\Sigma &= \int_0^\tau dt \int_0^\tau ds \int d\bm{x} \int d\bm{y} \ \big(\bm{\nu}^\text{st,T}(\bm{x}) \cdot \bm{B}^{-1} \bm{\nu}^\text{st}(\bm{x}) \big) \big(\bm{\nu}^\text{st,T}(\bm{y}) \cdot \bm{B}^{-1} \bm{\nu}^\text{st}(\bm{y}) \big) \Big( p(\bm{x},t ; \bm{y},s) - p^\text{st}(\bm{x}) p^\text{st}(\bm{y}) \Big) + 2 \Delta S^\text{irr},  \nn
\text{Cov}_{\Sigma,Z} &= \int_0^\tau dt \int_0^\tau ds \int d\bm{x} \int d\bm{y} \ z(\bm{x}) \big(\bm{\nu}^\text{st,T}(\bm{y}) \cdot \bm{B}^{-1} \bm{\nu}^\text{st}(\bm{y}) \big) \Big( p(\bm{x},t ; \bm{y},s) - p^\text{st}(\bm{x}) p^\text{st}(\bm{y}) \Big)  .
\end{align}
With these expressions, it is obvious that choosing $Z = \bar{\Sigma}$, i.~e.,
\begin{align}
z(\bm{x}) = \bm{\nu}^\text{st,T}(\bm{x}) \bm{B}^{-1} \bm{\nu}^\text{st}(\bm{x}),
\end{align}
results in
\begin{align}
\text{Var}_\Sigma  \big(1 - {\chi_{\Sigma,\bar{\Sigma}}}^2 \big)  = \text{Var}_\Sigma - \frac{{\text{Cov}_{\Sigma,\bar{\Sigma}}}^2}{\text{Var}_{\bar{\Sigma}}} =  2 \Delta S^\text{irr} .
\end{align}
This choice thus realizes the equality in the CTUR \eqref{correlation-TUR}.

For the case of a pure jump process, we have for the variances and covariance,
\begin{align}
\text{Var}_J &= \int_0^\tau dt \int_0^t ds \sum_{i,j,k,l} \omega_{i j} \omega_{k l} W_{i j} W_{k l} \Big( p(j,t \vert k,s) p_l^\text{st} + p(l,t \vert i,s) p_j^\text{st} - 2 p_j^\text{st} p_l^\text{st} \Big) + \tau \sum_{i,j} \omega_{ij}^2 W_{i j} p_j^\text{st}, \\
\text{Var}_Z &= \int_0^\tau dt \int_0^t ds \sum_{j,l} z_j z_l \Big( p(j,t \vert l,s) p_l^\text{st} + p(l,t \vert j,s) p_j^\text{st} - 2 p_j^\text{st} p_l^\text{st} \Big), \nn
\text{Cov}_{J,Z} &= \int_0^\tau dt \int_0^t ds \ \sum_{j,k,l} z_j \omega_{k l} W_{k l} \Big( p(j,t \vert k,s) p_l^\text{st} + p(l,t \vert j,s) p_j^\text{st} - 2 p_j^\text{st} p_l^\text{st} \Big) \n .
\end{align}
We decompose the transition rates into the (irreversible) local mean velocity and a reversible part,
\begin{align}
W_{i j} = V^\text{st}_{i j} + \Phi^\text{st}_{i j} \qquad \text{with} \qquad V_{ij}^\text{st} = \frac{1}{2} \bigg( W_{ij} - W_{ji} \frac{p_i^\text{st}}{p_j^\text{st}} \bigg) \qquad \text{and} \qquad \Phi_{ij}^\text{st} = \frac{1}{2} \bigg( W_{ij} + W_{ji} \frac{p_i^\text{st}}{p_j^\text{st}} \bigg) .
\end{align}
Using that $\omega_{ij} = \omega_{ji}$, we can write
\begin{align}
\text{Var}_J &= \int_0^\tau dt \int_0^t ds \sum_{i,j,k,l} \omega_{i j} \omega_{k l} \big(V^\text{st}_{i j} + \Phi^\text{st}_{i j} \big) \big(V^\text{st}_{k l} + \Phi^\text{st}_{k l} \big) \Big( p(j,t \vert k,s) p_l^\text{st} + p(l,t \vert i,s) p_j^\text{st} - 2 p_j^\text{st} p_l^\text{st} \Big) + \tau \sum_{i,j} \omega_{ij}^2 \Phi^\text{st}_{i j} p_j^\text{st}, \nn
\text{Cov}_{J,Z} &= \int_0^\tau dt \int_0^t ds \ \sum_{j,k,l} z_j \omega_{k l} \big(V^\text{st}_{k l} + \Phi^\text{st}_{k l} \big) \Big( p(j,t \vert k,s) p_l^\text{st} + p(l,t \vert j,s) p_j^\text{st} - 2 p_j^\text{st} p_l^\text{st} \Big). \label{current-variance-markov}
\end{align}
We focus on the stochastic pseudo-entropy $\mathcal{R}$ with the weighting function
\begin{align}
\omega_{i j} = 2 \frac{W_{ij} p_j^\text{st} - W_{j i} p_i^\text{st}}{W_{ij} p_j^\text{st} + W_{j i} p_i^\text{st}} = 2 \frac{V_{ij}^\text{st}}{\Phi_{ij}^\text{st}} .
\end{align}
For this choice, we obtain
\begin{align}
\sum_{k,l} \omega_{k l} \Phi_{kl}^\text{st} p(j,t \vert l,s) p_l^\text{st} = 2 \sum_{k,l} p(j,t \vert l,s) \big( W_{k l} p_l^\text{st} - W_{l k} p_k^\text{st} \big) = 2 \sum_l p(j,t \vert l,s) \sum_k \big( W_{k l} p_l^\text{st} - W_{l k} p_k^\text{st} \big) = 0,
\end{align}
where we used the steady state condition of the master equation.
As a consequence, all terms involving $\Phi_{ij}^\text{st}$ in \eqref{current-variance-markov} vanish, and we have
\begin{align}
\text{Var}_{\mathcal{R}} &= 4\int_0^\tau dt \int_0^t ds \sum_{i,j,k,l} \frac{(V_{ij}^\text{st})^2}{\Phi_{ij}^\text{st}} \frac{(V_{kl}^\text{st})^2}{\Phi_{kl}^\text{st}} \Big( p(j,t \vert l,s) p_l^\text{st} + p(l,t \vert j,s) p_j^\text{st} - 2 p_j^\text{st} p_l^\text{st} \Big) + 2 R ,\\
\text{Cov}_{\mathcal{R},Z} &= 2\int_0^\tau dt \int_0^t ds \ \sum_{j,k,l} z_j \frac{(V_{kl}^\text{st})^2}{\Phi_{kl}^\text{st}} \Big( p(j,t \vert l,s) p_l^\text{st} + p(l,t \vert j,s) p_j^\text{st} - 2 p_j^\text{st} p_l^\text{st} \Big). \n
\end{align}
Note that we can change the second argument of the conditional probabilities, since we have
\begin{align}
\sum_{i,j} \frac{(V_{ij}^\text{st})^2}{\Phi_{ij}^\text{st}} p(l,t \vert i,s) p_j^\text{st} = - \sum_{i,j} \frac{V_{ij}^\text{st}}{\Phi_{ij}^\text{st}} V_{ji}^\text{st} p(l,t \vert i,s) p_i^\text{st} = \sum_{i,j} \frac{V_{ji}^\text{st}}{\Phi_{ji}^\text{st}} V_{ji}^\text{st} p(l,t \vert i,s) p_i^\text{st} = \sum_{i,j} \frac{(V_{ij}^\text{st})^2}{\Phi_{ij}^\text{st}} p(l,t \vert j,s) p_j^\text{st},
\end{align}
where we used $V_{ij}^\text{st} p_j^\text{st} = - V_{ji}^\text{st} p_i^\text{st}$ in the first, $V_{ij}^\text{st}/\Phi_{ij}^\text{st} = - V_{ji}^\text{st}/\Phi_{ji}^\text{st}$ in the second and renamed indices in the third step.
We choose the function $z_j$ corresponding to the local average of the stochastic pseudo entropy $\bar{\mathcal{R}}$
\begin{align}
z_j = \sum_i \frac{\Big(W_{i j} - W_{j i} \frac{p_i^\text{st}}{p_j^\text{st}} \Big)^2}{W_{i j} + W_{j i} \frac{p_i^\text{st}}{p_j^\text{st}}} = 2 \sum_{i} \frac{(V_{ij}^\text{st})^2}{\Phi_{ij}^\text{st}},
\end{align}
which, when plugged into the above expressions, yields
\begin{align}
\text{Cov}_{\mathcal{R},\bar{\mathcal{R}}} = \text{Var}_{\bar{\mathcal{R}}} =  4\int_0^\tau dt \int_0^t ds \sum_{i,j,k,l} \frac{(V_{ij}^\text{st})^2}{\Phi_{ij}^\text{st}} \frac{(V_{kl}^\text{st})^2}{\Phi_{kl}^\text{st}} \Big( p(j,t \vert l,s) p_l^\text{st} + p(l,t \vert j,s) p_j^\text{st} - 2 p_j^\text{st} p_l^\text{st} \Big).
\end{align}
This is precisely the same as the first term in $\text{Var}_{\mathcal{R}}$ and thus, we finally obtain
\begin{align}
\text{Var}_{\mathcal{R}} \big( 1 - {\chi_{\mathcal{R},\bar{\mathcal{R}}}}^2 \big) = 2 R,
\end{align}
which yields equality in \eqref{TUR-corr-markov}.

\section{Dependence of the CTUR on the size of the data set and sampling interval} \label{app-error}
\begin{figure*}
\includegraphics[width=.49\textwidth]{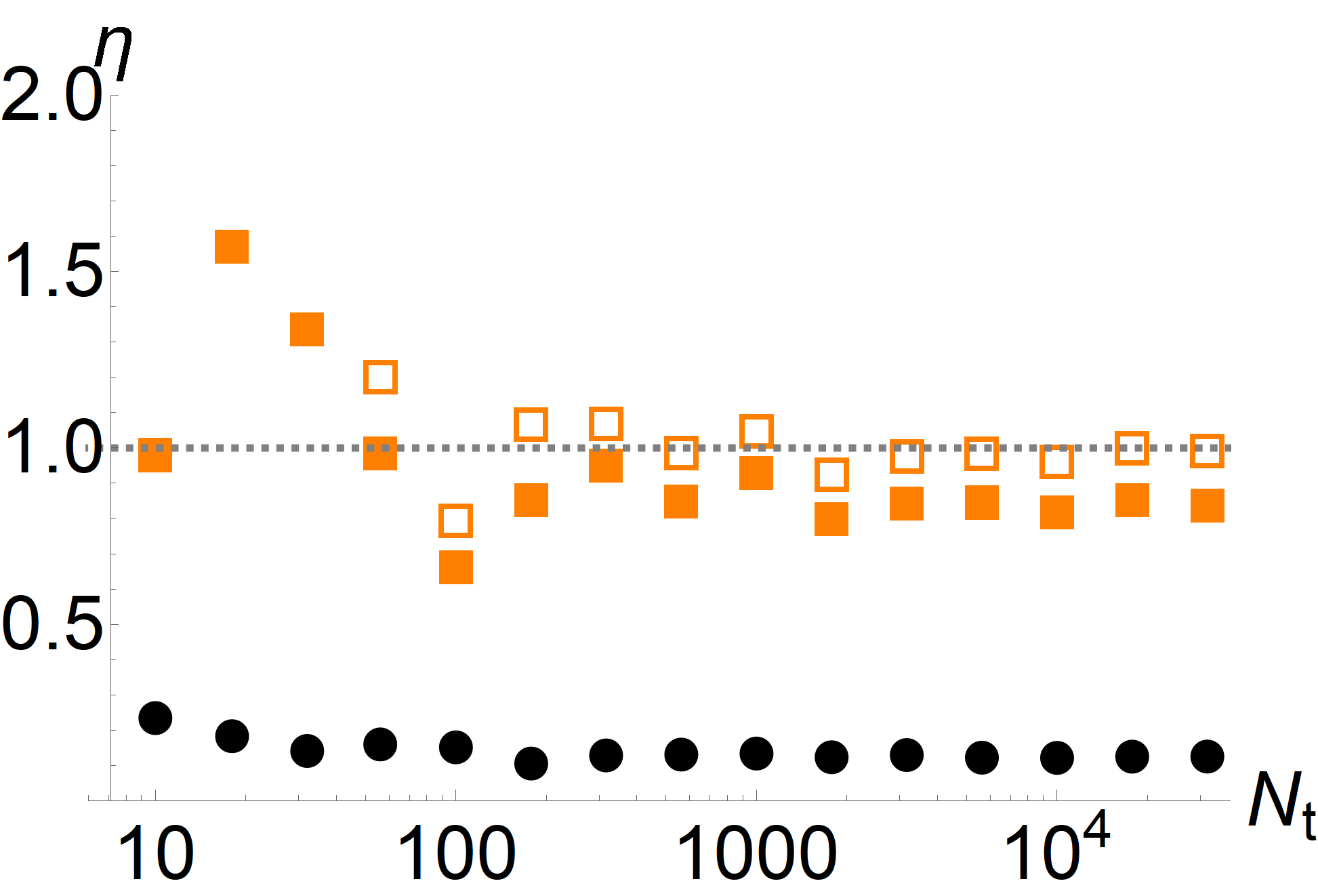}
\includegraphics[width=.49\textwidth]{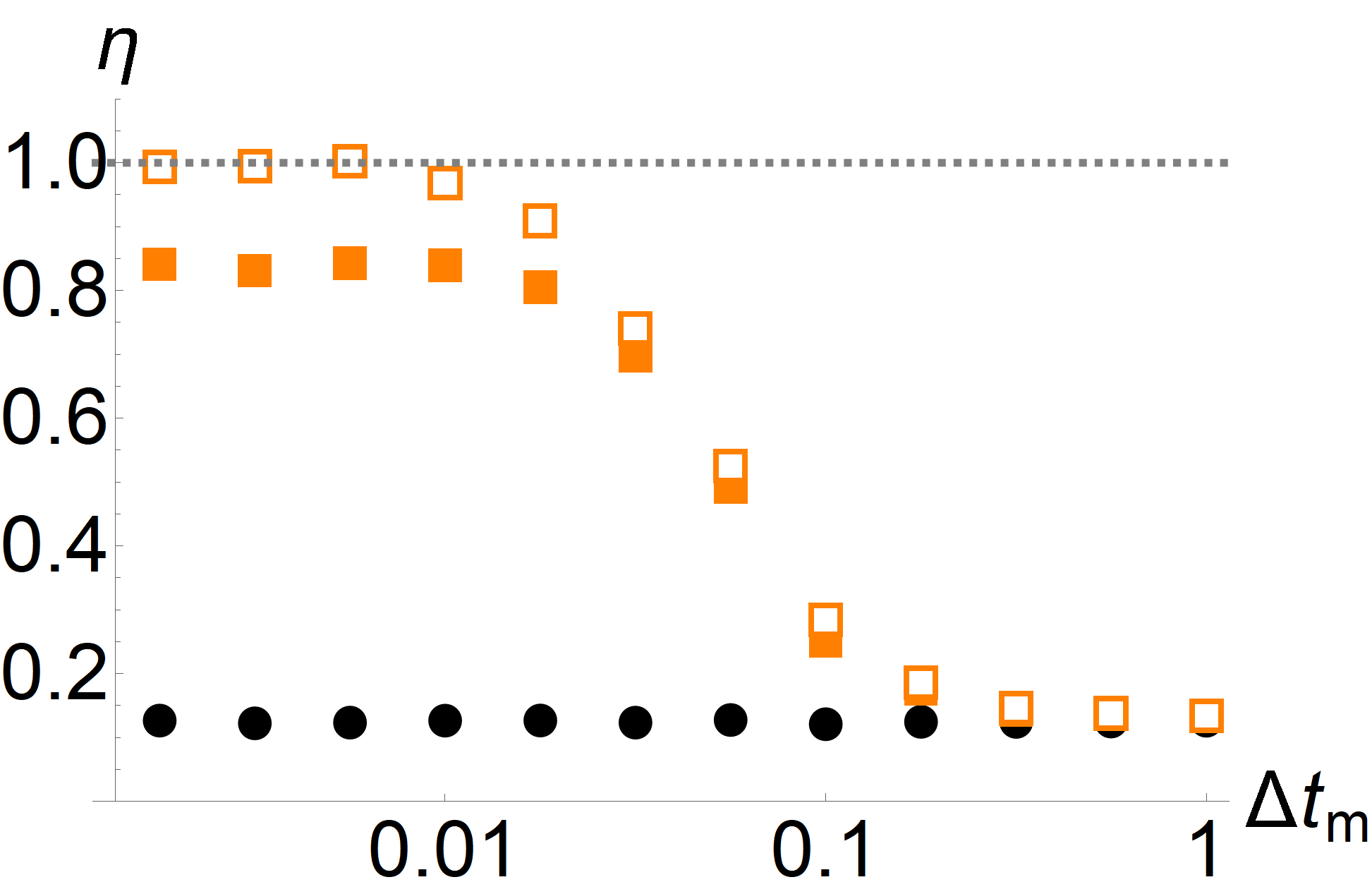}
\caption{Different estimates on the entropy production as a function of various simulation parameters for the particle in a tilted periodic potential.
In all panels, the black dots correspond to the ratio $\eta_J$ (see \eqref{transport-eff}) for the TUR, the filled orange squares to the ratio $\eta_{J,Z}$ for the CTUR with the observable $Z$ being the integral over the inverse occupation fraction and the empty orange squares to the ratio $\eta_{J,\tilde{Z}}$ with the numerically optimized observable $\tilde{Z}$. The parameter values are $U_0 = 1$, $L = 1$, $F_0 = 5$, $\mu = 1$ and $T = 0.2$; the length of the observation interval is $\tau = 100$. The left panel show the ratios as a function of the number of trajectories for a sampling interval $\Delta t_\text{m} = 0.01$, the right panel as a function of the sampling interval for $N_\text{t} = 5000$ trajectories.} \label{fig-error}
\end{figure*}

The CTUR \eqref{correlation-TUR} is a relation between the averages and variances of different physical observables.
Formally, these quantities are defined for an infinite ensemble.
However, in reality the number of trajectories $N_\text{t}$ is always finite, both in experiments and numerical simulations.
It is clear that the averages and variance computed from a finite number of random trajectories are themselves random quantities; thus, for any finite data set, there is a finite probability that the relation \eqref{correlation-TUR} can be violated.
Moreover, the state-dependent observable \eqref{nc-observable} is defined as a time-integral over continuous-time trajectories.
However, in reality, the sampling interval is likewise finite, that is, instead of a continuous trajectory $\lbrace \bm{x}(t) \rbrace_{t \in [0,\tau]}$, we observe a sequence of $M$ discrete values $\bm{x}(t_j)$ with $t_j = j \Delta t_\text{m}$ and $\Delta t^\text{m} = \tau/M$.
Here, we want to investigate the dependence of the bound \eqref{correlation-TUR} on $N_\text{t}$ and $\Delta t_\text{m}$.
For simplicity, we focus on a simple model that serves as a minimal example of a non-equilibrium steady state.
We consider a single overdamped particle in a one-dimensional periodic potential $U(x + L) = U(x)$, which is driven by a constant force $F_0$, while the environment is characterized by the mobility $\mu$ and temperature $T$.
The corresponding Langevin equation reads
\begin{align}
\dot{x}(t) = \mu \big( - U'(x) + F_0 \big) + \sqrt{2 \mu T} \xi(t) .
\end{align}
For long times, this system settles into a periodic steady state with probability density $p^\text{st}(x+L) = p^\text{st}(x)$.
Since the system is one-dimensional, this also fixes the steady-state local mean velocity $\nu^\text{st}(x) = v^\text{d}/(L p^\text{st}(x))$, see Section \ref{sec-data-occ}.
As a current observable, we consider the total displacement
\begin{align}
J = \int_0^\tau dt \ \dot{x}(t) .
\end{align}
As in Section \ref{sec-demonstration-motor}, we choose the state-dependent observable defined by the inverse probability density,
\begin{align}
Z = \int_0^\tau dt \ \frac{1}{p^\text{st}(x(t))},
\end{align}
which is proportional to the heuristic observable defined in Section \ref{sec-optimal-obs}.
$p^\text{st}(x)$ is estimated by the occupation fraction of the entire set of trajectories.
For a finite sampling interval $\Delta t_\text{m}$, we instead use the discrete approximation of the integral
\begin{align}
Z = \sum_{j=1}^M \frac{1}{p^\text{st}(x(t_j))} \Delta t_\text{m} \label{observabe-finite-interval}.
\end{align}
We further define an observable $\tilde{Z}$ parameterized as \eqref{parameter-observable} with $K = 10$ and numerically maximize the Pearson coefficient with respect to the parameters $a_k$ and $b_k$.
As a concrete example, we consider the case $U(x) = -U_0 \cos(2 \pi x/L)$ with $U_0 = 1$, $L = 1$, $F_0 = 5$, $\mu = 1$ and $T = 0.2$.
This corresponds to a moderately strong driving at relatively low temperature.
In this regime, the relevant timescale describing the dynamics is the time it takes the particle to traverse one period of the potential, $\tau^\text{d} = L/v^\text{d} \approx 1.3$.
For the above parameter values, the TUR is generally not tight; the ratio $\eta_J$ defined in \eqref{transport-eff} has a value of $\eta_J \approx 0.12$ and thus the ratio between the average current and its variance underestimates the entropy production by about a factor of $8$.
By contrast, as shown in Fig.~\ref{fig-error}, the CTUR with the numerically optimized observable $\tilde{Z}$ can provide an accurate estimate of the entropy production, given a sufficient amount of trajectories and a sufficiently small sampling interval.
As a function of $N_\text{t}$ (left panel of Fig.~\ref{fig-error}), the ratios corresponding to both the TUR and the CTUR show noticeable fluctuations for a small number of trajectories.
However, while the value of the TUR ratio stabilizes at around $N_\text{t} \approx 100$, we need around $N_\text{t} \approx 1000$ trajectories for the CTUR to provide a reliable estimate.
The reason is that, since, in this case, the CTUR is much tighter than the TUR, the Pearson coefficient $\chi_{J,Z}$ is close to $1$.
Since the left-hand side of \eqref{correlation-TUR} diverges as $\chi_{J,Z}$ approaches unity, small fluctuations in its value have a large impact on the estimate.
As a function of the sampling interval $\Delta t_\text{m}$, we note that, if $\Delta t_\text{m}$ is comparable to the transport timescale $\tau^\text{d}$, the CTUR does not yield any improvement over the TUR.
The reason is that, for such a large sampling interval, the spatially periodic function $p^\text{st}(x)$ in \eqref{observabe-finite-interval} can no longer resolve the motion of the particles.
Thus, the observable $Z$ is no longer correlated with the displacement of the particle, causing the Pearson coefficient to vanish.
Note that the TUR, which only depends on the accumulated current until the final time, is independent of the sampling interval.
We see that the CTUR can yield a significant improvement over the TUR for $\Delta t_\text{m} < 0.1 \tau^\text{d}$ and saturates at $\Delta t_\text{m} \approx 0.01 \tau^\text{d}$.
We remark that this is still two orders of magnitude larger than the sampling interval needed to accurately determine the entropy production directly from the trajectory using its representation as a stochastic current
\begin{align}
\Sigma = \frac{1}{T} \int_0^\tau dt \ F(x(t)) \circ \dot{x}(t) .
\end{align}

\end{document}